\pdfoutput=1
\RequirePackage{ifpdf}
\ifpdf 
\documentclass[pdftex]{sigma}
\else
\documentclass{sigma}
\fi

\usepackage{bm,bbm}
\usepackage{physics}
\usepackage[mathscr]{euscript}
\usepackage{tikz}
\usetikzlibrary{calc}
\usetikzlibrary{cd}
\usepackage[root radius=.8mm,edge length= 5mm,mark=o]{dynkin-diagrams}
\usepackage{enumitem}
\usepackage{accents}

\usepackage{booktabs}

\numberwithin{equation}{section}

\begin{document}
\allowdisplaybreaks

\newcommand{\arXivNumber}{2212.03870}

\renewcommand{\PaperNumber}{039}

\FirstPageHeading

\ShortArticleName{Double Quiver Gauge Theory and BPS/CFT Correspondence}

\ArticleName{Double Quiver Gauge Theory\\ and BPS/CFT Correspondence}

\Author{Taro KIMURA}

\AuthorNameForHeading{T.~Kimura}

\Address{Institut de Math\'ematiques de Bourgogne, Universit\'e de Bourgogne, CNRS, France}
\Email{\href{mailto:taro.kimura@u-bourgogne.fr}{taro.kimura@u-bourgogne.fr}}
\URLaddress{\url{https://kimura.pages.math.cnrs.fr/}}

\ArticleDates{Received January 22, 2023, in final form May 28, 2023; Published online June 08, 2023}

\Abstract{We provide a formalism using the $q$-Cartan matrix to compute the instanton partition function of quiver gauge theory on various manifolds. Applying this formalism to eight dimensional setups, we introduce the notion of double quiver gauge theory characterized by a pair of quivers. We also explore the BPS/CFT correspondence in eight dimensions based on the $q$-Cartan matrix formalism.}

\Keywords{quiver gauge theory; BPS/CFT correspondence; instanton moduli space; quiver variety; Calabi--Yau four-fold}

\Classification{81T30; 16G20; 14D21; 14J35}

\section{Introduction}

Towards profound understanding of quantum field theory, how to exactly evaluate physical observables is a ubiquitous problem.
The instanton partition function is one of the canonical examples of such an exact analysis, which has been playing a central role in the study of non-perturbative aspects of supersymmetric gauge theory~\cite{Losev:1997tp,Lossev:1997bz,Moore:1997dj,Nekrasov:2002qd}.
A purpose of this paper is to provide an alternative algebraic framework to obtain the instanton partition function for a~wide class of gauge theories.

Quiver gauge theory is a class of gauge theory, which involves multiple gauge degrees of freedom with a diagrammatic characterization by a quiver diagram.
In fact, quiver generalization of the instanton partition function exhibits various connections with interesting mathematical concepts~\cite{Nekrasov:2012xe,Nekrasov:2013xda}.
Another situation that benefits from a quiver description is the instanton on the asymptotically locally Euclidean (ALE) space obtained from the resolution of the orbifold singularity $\mathbb{C}^2/\Upsilon$ for $\Upsilon$ a finite subgroup of SU(2)~\cite{Kronheimer:1990MA}.
There is the so-called ${\rm ADE}$ classification of $\Upsilon$, and it has been established that the instanton moduli space of the ALE space is given by the quiver variety associated with the affine quiver of type ${\rm ADE}$ through the McKay--Nakajima correspondence~\cite{Nakajima:1994qu,Nakajima:1998DM}.
Observing this correspondence, geometric realization of quiver gauge theories has been established for various situations~\cite{Douglas:1996sw}.

\subsubsection*{Double quiver theory}

A key observation presented in this paper is that the orbifold instanton partition function of type $A$~\cite{Fucito:2004ry,Fujii:2005dk} is concisely formulated using the $q$-deformation of Cartan matrix associated with the corresponding affine quiver~\cite{Kimura:2015rgi}.
An important property of the $q$-deformed Cartan matrix is its invertibility: The Cartan matrix of affine quiver is not invertible before the $q$-deformation in general.
From this point of view, the standard Euclidean case $\mathbb{C}^2$ corresponds to $\widehat{A}_0$ quiver (Jordan quiver), where the instanton moduli space is given by the quiver variety associated with $\widehat{A}_0$ quiver, and the $q$-Cartan matrix of type $\widehat{A}_0$ plays a role there (maybe without noticing before).

In this paper, combining two formalisms based on quiver description, we propose the notion of double quiver gauge theory characterized by a pair of quivers.
The first quiver $\Gamma$ is to specify the multiple gauge degrees of freedom and their interaction as a quiver gauge theory.
The second quiver $\Upsilon$ is to characterize the manifold on which the gauge theory is defined.
In this formalism, the second quiver is not necessarily an affine quiver, but can be taken to be a generic quiver even though there is no direct geometric realization.
Moreover, we can even go beyond the standard simply-laced quivers by using the formalism of fractional quivers~\cite{Kimura:2017hez,KPfractional,Nakajima:2019olw}.
For example, for $\Upsilon = A_1$, the corresponding instanton partition function computes the equivariant volume of the cotangent bundle of the Grassmannian, which is obtained as the quiver variety of type $A_1$.
We may find a geometric origin of double quiver theory in the eight-dimensional setup, called the gauge origami~\cite{Nekrasov:2015wsu,Nekrasov:2016ydq}.
If both quivers are of affine type, it has a geometric realization $\big(\mathbb{C}^2/\Upsilon\big) \times \big(\mathbb{C}^2/\Gamma\big)$ with a proper choice of the framing space (the configuration of D3 branes in the string theory setup).
From this point of view, the $q$-Cartan matrix encodes the geometric information through the MaKay--Nakajima correspondence together with the $\Omega$-background dependence.

The formalism of double quiver theory is also applied to another eight-dimensional theory, called the magnificent four, which is realized as a D$(-1)$-D7-$\overline{\text{D7}}$ system in the IIB string theory (D0-D8-$\overline{\text{D8}}$ system in the IIA theory)~\cite{Nekrasov:2017cih,Nekrasov:2018xsb}.
In this paper, we obtain the contour integral form of the instanton partition function in eight dimensions with various orbifold structures based on the $q$-deformed Cartan matrix formalism.
Although we should take into account the Calabi--Yau condition for the four-fold, the construction is almost parallel with the double quiver theory up to the choice of the framing space.

\subsubsection*{BPS/CFT correspondence}

Another important concept concerning the supersymmetric gauge theory is the BPS/CFT correspondence, which is the correspondence between the BPS observables in SUSY gauge theory and the vertex operators in the CFT-like theories~\cite{Nekrasov:2004UA,Nekrasov:2015wsu}.
A canonical example is the equivalence between the instanton partition function on $\mathbb{C}^2$ and the Liouville/Toda conformal block, a.k.a., AGT-W relation~\cite{Alday:2009aq,Wyllard:2009hg}.
Along this direction, it has been pointed out that the contour integral form of the instanton partition function also has a realization using the vertex operators~\cite{Kimura:2019hnw}.
In this paper, we construct a formalism of vertex operators depending on a pair of quivers, where the geometric information is algebraically encoded in the $q$-Cartan matrix.
We demonstrate that the instanton partition function of double quiver theory and the magnificent four are naturally realized as a correlation function of these vertex operators.

\subsubsection*{Organization of the paper}

The remaining part of the paper is organized as follows:
After summarising the notations used in this paper in Section~\ref{sec:notation}, we in Section~\ref{sec:inst_part_fn} study the instanton partition function for quiver gauge theory and for the orbifold.
We in particular show that the $q$-deformed Cartan matrix plays a crucial role in the construction in these cases.
Based on this formalism, in Section~\ref{sec:doubld_quiver_theory}, we propose the notion of double quiver theory parametrized by a pair of quivers and show several fundamental examples.
In Section~\ref{sec:mag_four}, we study the eight-dimensional theory called the magnificent four on several backgrounds.
We show that the eight-dimensional instanton partition function is concisely obtained by applying the $q$-Cartan matrix formalism.
In Section~\ref{sec:BPS/CFT_corresp}, we study the BPS/CFT correspondence for double quiver theory and magnificent four.
Introducing the vertex operators depending on a pair of quivers, we see that the instanton partition function is realized as a correlation function of these vertex operators.

\section{Preliminaries}\label{sec:notation}

In this section, we summarize the notations used in this paper.

\subsubsection*{Special functions}

We use the $q$-factorial ($q$-Pochhammer) symbol
\begin{gather*}
 (x;q)_n = \prod_{m=0}^{n-1} (1 - x q^m) .
\end{gather*}
The theta function with the nome $p \in \mathbb{C}^\times$ is given by
\begin{gather*}
 \theta(x;p) = (x;p)_\infty (p/x;p)_\infty.
\end{gather*}

We define the $q$-deformed multiple gamma function for $|z| < 1$, $|q_i| < 1$ as follows:
\begin{align}
 \Gamma_k(z;q_1,\ldots,q_k)
 & = \prod_{0 \le n_1,\ldots,n_k \le \infty} \big(1 - z q_1^{n_1} \cdots q_k^{n_k}\big)^{(-1)^k}
 \nonumber \\
 & = \exp \left( (-1)^{k+1} \sum_{n=1}^\infty \frac{z^n}{n} \frac{1}{\big(1 - q_1^n\big) \cdots \big(1 - q_k^n\big)} \right).
 \label{eq:mult_gamma}
\end{align}
The elliptic gamma function is defined using the $q$-double gamma functions
\begin{gather}
 \Gamma_{\rm e}(e;q_1,q_2) = \frac{\Gamma_2\big(z^{-1} q_1 q_2 ;q_1,q_2\big)}{\Gamma_2(z;q_1,q_2)} . \label{eq:e_gamma}
\end{gather}

\subsubsection*{Index functor}

For the vector bundle with the virtual character
\begin{gather}
 \operatorname{ch} \mathbf{X} = \sum_{i=1} n_i {\rm e}^{x_i} , \label{eq:X-bundle}
\end{gather}
with $n_i \in \mathbb{Z}$ the multiplicity of the Chern root $x_i$, we define the index functor, converting the additive class to the multiplicative class,
\begin{gather*}
 \mathbb{I}[\mathbf{X}] = \prod_{i=1} [x_i]^{n_i} ,
\end{gather*}
where the bracket symbol is defined by
\begin{gather*}
 [x] =
 \begin{cases}
 x, & (4d) \\
 1 - {\rm e}^{-x}, & (5d) \\
 \theta\big({\rm e}^{-x};p\big). & (6d)
 \end{cases}
\end{gather*}
There is a hierarchical structure between rational, trigonometric, and elliptic functions,
\[
 \theta\big({\rm e}^{-x};p\big) \xrightarrow{p \to 0} 1 - {\rm e}^{-x} = x + O\big(x^2\big) .
\]
We remark the reflection formula
\[
 [-x] =
 \begin{cases}
 -[x], & (4d) \\
 - {\rm e}^x [x]. & (5d \ \& \ 6d)
 \end{cases}
\]
For the dual bundle having the character $\operatorname{ch} \mathbf{X}^\vee = \sum_{i = 1} n_i {\rm e}^{-x_i}$, we have
\[
 \mathbb{I}[\mathbf{X}^\vee]
 =
 \begin{cases}
 (-1)^{\operatorname{rk} \mathbf{X}} \mathbb{I}[\mathbf{X}], & (4d) \\
 (-1)^{\operatorname{rk} \mathbf{X}} \det \mathbf{X} \mathbb{I}[\mathbf{X}], & (5d \ \& \ 6d)
 \end{cases}
\]
where we define $\operatorname{rk} \mathbf{X} = \sum_i n_i$ and $\det \mathbf{X} = \prod_i {\rm e}^{n_i x_i}$.

The 4d theory index corresponds to the (equivariant) Euler class.
The 5d theory corresponds to K-theory convention, where the index is given by the character of the alternating sum of the dual bundle,
\[
 \mathbb{I}[\mathbf{X}] = \operatorname{ch} \wedge \mathbf{X}^\vee ,
\]
and, denoting $\operatorname{ch} \mathbf{y} = \mathsf{y}$, we have
\[
 \mathbb{I}[\mathbf{y}^\vee \mathbf{X}] = \operatorname{ch} \wedge_\mathsf{y} \mathbf{X}^\vee,
\]
where we denote the alternating sum of the wedge products by
\[
 \wedge \mathbf{X} = \sum_{i = 0}^{\operatorname{rk} \mathbf{X}} (-1)^i \wedge^i \mathbf{X} , \qquad
 \wedge_y \mathbf{X} = \sum_{i = 0}^{\operatorname{rk} \mathbf{X}} (-y)^i \wedge^i \mathbf{X} .
\]

In addition, we denote the vector bundle constructed from $\mathbf{X}$ via the $p$-th Adams operation by $\mathbf{X}^{[p]}$.
Given the original character \eqref{eq:X-bundle}, the character is given by
\begin{gather}
 \operatorname{ch} \mathbf{X}^{[p]} = \sum_{i=1}^{\operatorname{rk} \mathbf{X}} {\rm e}^{p x_i} . \label{eq:Adams_op}
\end{gather}
Hence, the dual bundle is formally written as $\mathbf{X}^\vee = \mathbf{X}^{[-1]}$.

\subsubsection*{Characteristic polynomial}

Denoting a one-dimensional bundle $\mathbf{y}$ with $\operatorname{ch} \mathbf{y} = {\rm e}^y$, we define the characteristic polynomial with respect to the bundle $\mathbf{X}$,
\begin{gather*}
 P_\mathbf{X}(y) = \mathbb{I}\big[\mathbf{X}^\vee \mathbf{y}\big] = \prod_{i = 1}^{\operatorname{rk} \mathbf{X}} [y - x_i] , \qquad
 \widetilde{P}_\mathbf{X}(y) = \mathbb{I}\big[\mathbf{y}^\vee \mathbf{X}\big] = \prod_{i = 1}^{\operatorname{rk} \mathbf{X}} [x_i - y] ,
\end{gather*}
with the relation
\[
 \widetilde{P}_\mathbf{X}(y) =
 \begin{cases}
 (-1)^{\operatorname{rk} \mathbf{X}} P_\mathbf{X}(y), & (4d) \\
 (-{\rm e}^{y})^{\operatorname{rk} \mathbf{X}} \det \mathbf{X}^\vee P_\mathbf{X}(y). & (5d \ \& \ 6d)
 \end{cases}
\]

\subsubsection*{$\boldsymbol{\mathscr{S}}$-function}

For the parameters $(\epsilon_{1},\epsilon_2)$, we define the $\mathscr{S}$-function,
\begin{gather}
 \mathscr{S}(x) = \frac{[x - \epsilon_{1,2}]}{[x][x - \epsilon_{12}]}, \label{eq:S-func}
\end{gather}
where we use the notation
\[
 [x - \epsilon_{1,2}] = [x - \epsilon_1][x - \epsilon_2].
\]
We remark the reflection formula
\[
 \mathscr{S}(x) = \mathscr{S}(- x + \epsilon_{12}).
\]

When we have more parameters, we also define a generalized $\mathscr{S}$-function,
\[
 \mathscr{S}_{ij}(x) = \frac{[x-\epsilon_{i,j}]}{[x][x-\epsilon_{ij}]}
\]
where we denote $\epsilon_{ij} = \epsilon_i + \epsilon_j$ and $[x - \epsilon_{i,j}] = [x - \epsilon_i][x - \epsilon_j]$.
We also use the notation, $\epsilon_{i^a j^b} = a \epsilon_i + b \epsilon_j$.
The reflection formula is in this case given by
\[
 \mathscr{S}_{ij}(x) = \mathscr{S}_{ij}(- x + \epsilon_{ij}).
\]

\section{Instanton partition function}\label{sec:inst_part_fn}

\subsection{Pure gauge theory}

We consider the $k$-instanton moduli space of U($n$) gauge theory on $\mathcal{S} = \mathbb{C}^2$ denoted by $\mathfrak{M}_\gamma$ with the topological data $\gamma = (n,k)$.
We introduce the vector spaces
\begin{gather}
 \mathbf{N} = \mathbb{C}^n , \qquad \mathbf{K} = \mathbb{C}^k , \label{eq:N_K_vect_sp}
\end{gather}
which we call the framing space and the instanton space.
We have the automorphism groups, $\mathrm{GL}(\mathbf{N})$ and $\mathrm{GL}(\mathbf{K})$, and the characters
\[
 \operatorname{ch} \mathbf{N} = \sum_{\alpha = 1}^n {\rm e}^{a_\alpha} , \qquad
 \operatorname{ch} \mathbf{K} = \sum_{I = 1}^k {\rm e}^{\phi_I} .
\]
We denote the fiber of the cotangent bundle at the marked point $o \in \mathcal{S}$ by
\[
 \mathbf{Q} = T_o^\vee \mathcal{S} = \mathbf{Q}_1 \oplus \mathbf{Q}_2
\]
with the character
\[
 \operatorname{ch} \mathbf{Q}_i = {\rm e}^{\epsilon_i} = q_i, \qquad i = 1,2 .
\]
The marked point is now taken to be the fixed point under the $\mathrm{GL}(\mathbf{Q})$ action, i.e., $o = (0,0) \in \mathcal{S}$.
The parameters $(a_\alpha,\phi_I,\epsilon_{1,2})$ are the equivariant parameters of the corresponding group actions, $(\mathrm{GL}(\mathbf{N}), \mathrm{GL}(\mathbf{K}), \mathrm{GL}(\mathbf{\mathbf{Q}}))$, respectively:
These parameters are the elements of the corresponding Cartan subalgebra, $(a_\alpha)_{\alpha =1, \ldots,n} \in \operatorname{Lie} \mathbb{T}_\mathbf{N}$, $(\phi_I)_{I =1, \ldots,k} \in \operatorname{Lie} \mathbb{T}_\mathbf{K}$, and $(\epsilon_{1,2}) \in \operatorname{Lie} \mathbb{T}_\mathbf{Q}$ where we denote the Cartan tori by $\mathbb{T}_\mathbf{N} \subset \mathrm{GL}(\mathbf{N})$, $\mathbb{T}_\mathbf{K} \subset \mathrm{GL}(\mathbf{K})$, and $\mathbb{T}_\mathbf{Q} \subset \mathrm{GL}(\mathbf{Q})$.

We use the notation
\begin{subequations}\label{eq:P_Q_notation}
\begin{gather}
 \mathbf{P}_i = \wedge \mathbf{Q}_i = 1 - \mathbf{Q}_i, \qquad i = 1,2, \\
 \mathbf{P}_{12} = \wedge \mathbf{Q} = 1 - \mathbf{Q}_1 - \mathbf{Q}_2 + \mathbf{Q}_{12} , \qquad
 \mathbf{Q}_{12} = \wedge^2 \mathbf{Q} = \det \mathbf{Q} = \mathbf{Q}_1 \mathbf{Q}_2 .
\end{gather}
\end{subequations}
We remark the factorization property of $\mathbf{P}_{12}$: $\mathbf{P}_{12} = \mathbf{P}_1 \mathbf{P}_2$, and we will also use the notation (Section~\ref{sec:mag_four}),
\[
 \mathbf{P}_{123} = \mathbf{P}_1 \mathbf{P}_2 \mathbf{P}_3 , \qquad
 \mathbf{P}_{1234} = \mathbf{P}_1 \mathbf{P}_2 \mathbf{P}_3 \mathbf{P}_4 , \qquad
 \mathbf{Q}_{1234} = \mathbf{Q}_1 \mathbf{Q}_2 \mathbf{Q}_3 \mathbf{Q}_4 .
\]
Then, we have the observable sheaf associated with a marked point $o \in \mathcal{S}$ in K-theory as follows:
\[
 \mathbf{Y}_o \equiv \mathbf{Y} = \mathbf{N} - \mathbf{P}_{12} \mathbf{K},
\]
such that the universal sheaf $\mathbf{Y}_\mathcal{S}$ on $\mathfrak{M}_{n,k} \times \mathcal{S}$ is given by localization via the inclusion $i_o\colon o \hookrightarrow \mathcal{S}$
\[
 \mathbf{Y}_\mathcal{S} = \frac{\mathbf{Y}_o}{\mathbf{P}_{12}}.
\]

The vector multiplet contribution, which is associated with the tangent bundle to the moduli space $\mathfrak{M}_\gamma$, is constructed from the observable sheaf,
\[
 \mathbf{V} = \frac{\mathbf{Y}^\vee \mathbf{Y}}{\mathbf{P}_{12}}
 = \frac{\mathbf{N}^\vee \mathbf{N}}{\mathbf{P}_{12}} - \mathbf{N}^\vee \mathbf{K} - \mathbf{Q}_{12}^\vee \mathbf{K}^\vee \mathbf{N} + \mathbf{P}_{12}^\vee \mathbf{K}^\vee \mathbf{K}
 = \mathring{\mathbf{V}} + \mathbf{V}_\text{inst},
\]
where we split it into the perturbative part and the instanton part,
\[
 \mathring{\mathbf{V}} = \frac{\mathbf{N}^\vee \mathbf{N}}{\mathbf{P}_{12}} , \qquad
 \mathbf{V}_\text{inst} = - \mathbf{N}^\vee \mathbf{K} - \mathbf{Q}_{12}^\vee \mathbf{K}^\vee \mathbf{N} + \mathbf{P}_{12}^\vee \mathbf{K}^\vee \mathbf{K} .
\]
In general, the instanton part of the vector bundle $\mathbf{X}$ is given by $\mathbf{X}_\text{inst} = \mathbf{X} - \mathring{\mathbf{X}}$, where we define the perturbative part of $\mathbf{X}$ by $\mathring{\mathbf{X}} = \mathbf{X}\big|_{\mathbf{K} \to 0}$.

\subsubsection{Contour integral formula}

Let $\mathbf{x}$ be a one-dimensional bundle with the Chern root $x$.
Then, we define the gauge polynomial as the characteristic polynomial with respect to the vector space $\mathbf{N}$,
\begin{gather}
 P_\mathbf{N}(x) = \mathbb{I}\big[\mathbf{N}^\vee \mathbf{x}\big] = \prod_{\alpha = 1}^n [x - a_\alpha] , \qquad
 \widetilde{P}_\mathbf{N}(x) = \mathbb{I}\big[\mathbf{x}^\vee \mathbf{N}\big]
 = \prod_{\alpha = 1}^n [- x + a_\alpha] \label{eq:gauge_poly}
\end{gather}
with the relation
\begin{gather}
 \widetilde{P}_\mathbf{N}(x) =
 \begin{cases}
 (-1)^n P_\mathbf{N}(x), & (4d) \\
 \displaystyle
 (-1)^n {\rm e}^{n x} {\rm e}^{- a} P_\mathbf{N}(x), & (5d \ \& \ 6d)
 \end{cases}
 \label{eq:gauge_poly_reflection}
\end{gather}
where we denote $a = \sum_{\alpha = 1}^n a_\alpha$.
Hence, the special unitary condition is given by $a = 0$.

Applying the index functor to the vector multiplet contribution, we obtain the instanton partition function for pure U($n$) gauge theory
\[
 Z_\gamma = \mathbb{I} [\mathbf{V}]_\text{inst} .
\]
This index formula yields the contour integral form for the instanton partition function, a.k.a., LMNS formula~\cite{Losev:1997tp,Lossev:1997bz,Moore:1997dj,Nekrasov:2002qd}
\begin{gather}
 Z_\gamma = \frac{1}{k!} \frac{[-\epsilon_{12}]^k}{[-\epsilon_{1,2}]^{k}} \oint_{\mathbb{T}_\mathbf{K}} \dd{\underline{\phi}} \prod_{I=1}^k \frac{1}{P_\mathbf{N}(\phi_I) \widetilde{P}_\mathbf{N}(\phi_I + \epsilon_{12})} \prod_{I < J}^k \mathscr{S} (\phi_{IJ})^{-1} \mathscr{S} (\phi_{IJ} + \epsilon_{12})^{-1} .
 \label{eq:LMNS_formula}
\end{gather}
We denote $\dd{\underline{\phi}} = \prod_{I = 1}^k \dd{\phi}_I / 2 \pi \iota$ where $\iota^2 = -1$ and $\phi_{IJ} = \phi_I - \phi_J$.
The $\mathscr{S}$-function is defined in~\eqref{eq:S-func}.
The factor $k!$ is understood as the volume of Weyl group, $\operatorname{Weyl}(\mathrm{U}(k)) = \mathfrak{S}_k$.
From the analogy with the matrix model, we call the $\phi$-$\phi$ interaction part the measure term, and the characteristic polynomial part the source term.
We may also have the Chern--Simons factor $\prod_{I=1}^k {\rm e}^{\kappa \phi_I}$ in particular for 5d theory, which comes from the determinant bundle associated with~$\mathbf{Y}$.
The contour integral is taken over the Cartan torus of ${\rm GL}(\mathbf{K})$ denoted by ${\mathbb{T}_\mathbf{K}}$.
We remark that we have the zero mode factors in the character
\[
 \operatorname{ch} \mathbf{P}_{12}^\vee \mathbf{K}^\vee \mathbf{K}
 = \big(1 - q_1^{-1}\big)\big(1 - q_2^{-1}\big) \sum_{1 \le I \neq J \le k} {\rm e}^{\phi_{IJ}}
 + k (-q_1-q_2+q_{12}) + \underbrace{\sum_{I=1}^k 1}_{\text{zero modes}} ,
\]
and we interpret their index as the contour integral over the $\phi$-variable (integral over the Cartan torus $\mathbb{T}_\mathbf{K}$), which is equivalent to taking the residues.
Such a multivariable contour integral is organized by the Jeffrey--Kirwan residue prescription~\cite{Benini:2013xpa,Hori:2014tda,Hwang:2014uwa}.
See also \cite{Cordova:2014oxa}.

\subsubsection{Localization formula}

Applying the equivariant localization formula, the integral over the moduli space is localized on the fixed points under the equivariant action.
We introduce $n$-tuple partitions of size $k$,
\[
 \lambda = (\lambda_\alpha)_{\alpha = 1,\ldots,n} , \qquad
 |\lambda| = \sum_{\alpha = 1}^n |\lambda_\alpha| = k .
\]
Then, the equivariant fixed point in the instanton moduli space is given as follows:
\[
 \{ \phi_I \}_{I = 1,\ldots,k}
 \ \longrightarrow \
 \{ a_\alpha + (i-1) \epsilon_1 + (j-1) \epsilon_2 \}_{\alpha = 1,\ldots,n, (i,j) \in \lambda_\alpha} .
\]
This corresponds to the pole in the integrand of the contour integral formula of the instanton partition function.
Hence, the residue computation agrees with the localization formula.
The equivariant character of the vector space~$\mathbf{K}$ at the fixed point $\lambda$ is given by
\[
 \operatorname{ch} \mathbf{K}\big|_{\lambda} = \sum_{\alpha = 1}^n \sum_{(i,j) \in \lambda_\alpha} {\rm e}^{\alpha} q_1^{i-1} q_2^{j-1} .
\]
At the fixed point under the equivariant action, the observable sheaf is given by
\[
 \mathbf{Y} = \mathbf{P}_1 \mathbf{X},
\]
where we have
\[
 \operatorname{ch} \mathbf{X} = \sum_{x \in \mathcal{X}} x , \qquad
 \mathcal{X} = \big\{ {\rm e}^{a_\alpha} q_1^{k-1} q_2^{\lambda_{\alpha,k}} \big\}_{\alpha = 1,\ldots,n, k = 1,\ldots,\infty} .
\]
This expression corresponds to a partial reduction of the universal sheaf $\mathbf{Y}_\mathcal{S} = \mathbf{X} / \mathbf{P}_2$.
Exchanging $\epsilon_1 \leftrightarrow \epsilon_2$ $(q_1 \leftrightarrow q_2)$, we obtain another reduction $\mathbf{Y}_\mathcal{S} = \widetilde{\mathbf{X}} / \mathbf{P}_1$, where $\widetilde{\mathbf{X}}$ is obtained from transposed partitions.
We also define the configuration corresponding to $\lambda = \varnothing$,
\[
 \mathring{\mathcal{X}} = \big\{ {\rm e}^{a_\alpha} q_1^{k-1} \big\}_{\alpha = 1,\ldots,n, k = 1,\ldots,\infty} .
\]

We consider the vector multiplet contribution at the fixed point
\[
 \mathbf{V}\big|_{\mathcal{X}} = \frac{\mathbf{P}_1^\vee}{\mathbf{P}_2} \mathbf{X}^\vee \mathbf{X} .
\]
Denoting the set of fixed point configurations in the $k$-instanton sector by $\mathfrak{M}_{\gamma}^\mathsf{T} = \mathfrak{M}_{n,k}^\mathsf{T}$, we define their disjoint union, $\mathfrak{M}^\mathsf{T} = \bigsqcup_{k=0}^\infty \mathfrak{M}_{n,k}^\mathsf{T}$.
The full partition function evaluated at the fixed point configuration $\mathcal{X} \in \mathfrak{M}^\mathsf{T}$ is given by applying the index functor
\[
 Z_{\mathcal{X}} = \mathbb{I}[\mathbf{V}]_{\mathcal{X}} = \prod_{(x,x') \in \mathcal{X} \times \mathcal{X}} \frac{(q_{12} x/x';q_2)_\infty}{(q_2x/x';q_2)_\infty} ,
\]
and the total partition function is obtained by summing over all the fixed point configurations
\[
 Z = \sum_{\mathcal{X} \in \mathfrak{M}^\mathsf{T}} \mathfrak{q}^{k(\mathcal{X})} Z_{\mathcal{X}},
\]
where $\mathfrak{q}$ is the instanton counting parameter (fugacity; gauge coupling) and we denote the topological number associated with the configuration~$\mathcal{X}$ by $k(\mathcal{X}) = k$ for $\mathcal{X} \in \mathfrak{M}_{n,k}^\mathsf{T}$.

\subsubsection{Including a defect}\label{sec:defect}

We can extend the formalism of the equivariant integral presented in the previous part to the situation with a codimension-four defect~\cite{Agarwal:2018tso,Haouzi:2019jzk, Kim:2016qqs,Nekrasov:2015wsu, Tong:2014cha}.

In order to describe the defect, we first define the $\mathscr{Y}$-function and its dual~\cite{Nekrasov:2012xe,Nekrasov:2013xda}
\begin{gather*}
 \mathscr{Y}(x) = \mathbb{I}\big[\mathbf{Y}^\vee \mathbf{x}\big] = P_\mathbf{N}(x) \prod_{I = 1}^k \mathscr{S}(x - \phi_I) , \\
 \widetilde{\mathscr{Y}}(x) = \mathbb{I}\big[\mathbf{x}^\vee \mathbf{Y}\big] = \widetilde{P}_\mathbf{N}(x) \prod_{I = 1}^k \mathscr{S}(x - \phi_I) ,
\end{gather*}
where we denote a one-dimensional bundle $\mathbf{x}$ with the Chern root $x$ as before, and $P_\mathbf{N}(x)$ and $\widetilde{P}_\mathbf{N}(x)$ are the gauge polynomials defined in~\eqref{eq:gauge_poly}.
We have a similar relation between $\mathscr{Y}$ and $\widetilde{\mathscr{Y}}$ as in the case of the gauge polynomials~\eqref{eq:gauge_poly_reflection}.
Hereafter we will focus on $\mathscr{Y}(x)$.
We also use the notation $\mathscr{Y}_\mathbf{X} = \prod_{\alpha = 1}^{\operatorname{rk} \mathbf{X}} \mathscr{Y}(x_\alpha)$ for a generic bundle $\mathbf{X}$ with the Chern roots $(x_\alpha)_{\alpha=1,\ldots,\operatorname{rk}\mathbf{X}}$.

In addition to the vector spaces~\eqref{eq:N_K_vect_sp}, we introduce another vector space and with the character
\[
 \mathbf{N}' = \mathbb{C}^{{n}'} , \qquad
 \operatorname{ch} \mathbf{N}' = \sum_{\beta = 1}^{{n}'} {\rm e}^{b_\beta} ,
\]
which is interpreted as another framing space in the context of the instanton moduli space.
Denoting the topological data similarly by $\gamma = (n,{n}',k)$, we have the instanton partition function in the presence of the defect contribution
\begin{align*}
 Z_\gamma & = \frac{1}{k!} \frac{[-\epsilon_{12}]^k}{[-\epsilon_{1,2}]^{k}} \oint_{\mathbb{T}_K} \dd{\underline{\phi}} \mathscr{Y}_{\mathbf{N}'} \prod_{I=1}^k \frac{1}{P_\mathbf{N}(\phi_I) \widetilde{P}_\mathbf{N}(\phi_I + \epsilon_{12})} \prod_{I < J}^k \mathscr{S} (\phi_{IJ})^{-1} \mathscr{S} (\phi_{IJ} + \epsilon_{12})^{-1}
 \nonumber \\
 & =
 \frac{1}{k!} \frac{[-\epsilon_{12}]^k}{[-\epsilon_{1,2}]^{k}} \oint_{\mathbb{T}_K} \dd{\underline{\phi}} \frac{\prod^{\alpha = 1,\ldots,n}_{\beta = 1,\ldots,n'} [b_\beta - a_\alpha]}{\prod_{I = 1,\ldots,k}^{\alpha = 1,\ldots,n} [\phi_I - a_\alpha] [a_\alpha - \phi_I - \epsilon_{12}]} \frac{\prod_{I = 1,\ldots,k}^{\beta = 1,\ldots,n'} \mathscr{S}(b_\beta - \phi_I)}{\prod_{1 \le I < J \le k} \mathscr{S} (\phi_{IJ}) \mathscr{S} (\phi_{IJ} + \epsilon_{12})} .
\end{align*}
In this case, we have additional $\mathscr{S}$-functions in the integrand, which change the pole structure.
See~\cite{Agarwal:2018tso,Kim:2016qqs, Nekrasov:2015wsu} for details.
The instanton sum with this defect operator gives rise to the gauge theory average of the $qq$-character, which turns out to be a pole-free function of $b = \sum_{\beta=1}^{n'} b_\beta$~\cite{Nekrasov:2016qym}.

\subsection{Quiver gauge theory}\label{sec:quiver_gauge_theory}

We define a quiver as a set of nodes and (oriented) edges, $\Gamma = (\Gamma_0,\Gamma_1)$, where $\Gamma_0 = \{\text{nodes}\}$, $\Gamma_1 = \{ \text{edges} \}$.
As we consider quiver gauge theory with eight supercharges, all the edges have their dual edges with opposite orientation denoted by $\Gamma_1^\vee = \{e^\vee, e \in \Gamma_1 \}$.
We assign vector spaces for each node
\[
 \mathbf{N} = (\mathbf{N}_i)_{i \in \Gamma_0} , \qquad
 \mathbf{K} = (\mathbf{K}_i)_{i \in \Gamma_0},
\]
where
\[
 \mathbf{N}_i = \mathbb{C}^{n_i} , \qquad \mathbf{K}_i = \mathbb{C}^{k_i} .
\]
We denote the dimension vectors by $\gamma = (\underline{n},\underline{k}) = (n_i,k_i)_{i \in \Gamma_0}$.
The corresponding automorphism groups are given by
\[
 \mathrm{GL}(\mathbf{N}) = \prod_{i \in \Gamma_0} \mathrm{GL}(\mathbf{N}_i) , \qquad
 \mathrm{GL}(\mathbf{K}) = \prod_{i \in \Gamma_0} \mathrm{GL}(\mathbf{K}_i) ,
\]
and we have the characters
\[
 \operatorname{ch} \mathbf{N}_i = \sum_{\alpha = 1}^{n_i} {\rm e}^{a_{i,\alpha}} , \qquad
 \operatorname{ch} \mathbf{K}_i = \sum_{I = 1}^{k_i} {\rm e}^{\phi_{i,I}} .
\]
We similarly consider observable sheaves for each node
\[
 \mathbf{Y} = (\mathbf{Y}_i)_{i \in \Gamma_0} , \qquad
 \mathbf{Y}_i = \mathbf{N}_i - \mathbf{P}_{12} \mathbf{K}_i .
\]
In this case, we have the vector multiplet contribution assigned to the node $i \in \Gamma_0$ as follows:
\[
 \mathbf{V}_i = \frac{\mathbf{Y}_i^\vee \mathbf{Y}_i}{\mathbf{P}_{12}} = \frac{\mathbf{N}_i^\vee \mathbf{N}_i}{\mathbf{P}_{12}} - \mathbf{N}_i^\vee \mathbf{K}_i - \mathbf{Q}_{12}^\vee \mathbf{K}_i^\vee \mathbf{N}_i + \mathbf{P}_{12}^\vee \mathbf{K}_i^\vee \mathbf{K}_i .
\]

\subsubsection*{Bifundamental matter}

We introduce the bifundamental hypermultiplet contribution as follows:
\[
 \mathbf{H}_{e\colon i \to j} = - \mathbf{M}_e \frac{\mathbf{Y}_i^\vee \mathbf{Y}_j}{\mathbf{P}_{12}} = - \mathbf{M}_e \frac{\mathbf{N}_i^\vee \mathbf{N}_j}{\mathbf{P}_{12}} + \mathbf{M}_e \mathbf{N}_i^\vee \mathbf{K}_j + \mathbf{M}_e \mathbf{Q}_{12}^\vee \mathbf{K}_i^\vee \mathbf{N}_j - \mathbf{M}_e \mathbf{P}_{12}^\vee \mathbf{K}_i^\vee \mathbf{K}_j,
\]
where $\mathbf{M}_e$ is a one-dimensional bundle assigned to each edge $e \in \Gamma_1$.
The character is given by $\operatorname{ch} \mathbf{M}_e = {\rm e}^{m_e}$ with the bifundamental mass parameter $m_e$ identified with the Chern root.
We remark that all these bifundamental mass parameters can be fixed to be zero through gauge fixing except for cyclic quivers.
In particular, the instanton part of the vector and bifundamental hypermultiplet contribution is given by
\[
 \bigg[ \sum_{i \in \Gamma_0} \mathbf{V}_i + \sum_{e \in \Gamma_1} \mathbf{H}_e \bigg]_\text{inst}
 = - \sum_{i,j \in \Gamma_0} \big( \mathbf{N}_i^\vee \mathbf{c}_{ij}^{+\vee} \mathbf{K}_j + \mathbf{N}_i \mathbf{c}_{ij}^{-} \mathbf{K}_j^\vee \big) + \mathbf{P}_{12}^\vee \sum_{i,j \in \Gamma_0} \mathbf{K}_i^\vee \mathbf{c}_{ij}^{+\vee} \mathbf{K}_j
 ,
\]
where $\mathbf{c}^\pm$ are the half $q$-Cartan matrices associated with quiver $\Gamma$~\cite{Kimura:2015rgi}
\begin{gather}
 \mathbf{c}^+_{ij} = \delta_{i,j} - \sum_{e \in \Gamma_1\colon i \to j} \mathbf{M}_e^\vee , \qquad
 \mathbf{c}^-_{ij} = \mathbf{Q}_{12}^\vee \mathbf{c}^{+\vee}_{ji} .
 \label{eq:half_Cartan_mat}
\end{gather}
The total $q$-Cartan matrix is given by
\[
 \mathbf{c}_{ij} = \mathbf{c}^+_{ij} + \mathbf{c}^-_{ij} = \big(1 + \mathbf{Q}_{12}^\vee \big) \delta_{i,j} - \sum_{e \in \Gamma_1: i \to j} \mathbf{M}_e^\vee - \sum_{{\rm e}^\vee \in \Gamma_1^\vee: j \to i} \mathbf{M}_e \mathbf{Q}_{12}^\vee , \qquad i, j \in \Gamma_0 .
 \label{eq:tot_Cartan_mat}
\]
This $q$-Cartan matrix structure has a geometric origin. See Section~\ref{sec:geometry}.

\subsubsection*{Fundamental matter}

Similarly, we have the fundamental and antifundamental hypermultiplet contributions
\begin{gather*}
 \mathbf{H}_{i} = - \frac{\mathbf{M}_i^\vee \mathbf{Y}_i}{\mathbf{P}_{12}} = - \frac{\mathbf{M}_i^\vee \mathbf{N}_i}{\mathbf{P}_{12}} + \mathbf{M}_i \mathbf{K}_i , \\
 \widetilde{\mathbf{H}}_{i} = - \frac{\mathbf{Y}_i^\vee \widetilde{\mathbf{M}}_i}{\mathbf{P}_{12}} = - \frac{\mathbf{N}_i^\vee \widetilde{\mathbf{M}}_i}{\mathbf{P}_{12}} + \mathbf{Q}_{12}^\vee \mathbf{K}_i^\vee \widetilde{\mathbf{M}}_i,
\end{gather*}
where $\big(\mathbf{M},\widetilde{\mathbf{M}}\big) = \big(\mathbf{M}_i,\widetilde{\mathbf{M}}_i\big)_{i \in \Gamma_0}$ is the matter bundle.
Their Chern roots are identified with the (anti)fundamental mass parameters
\[
 \operatorname{ch} \mathbf{M}_i = \sum_{f = 1}^{n^\text{f}_i} {\rm e}^{m_{i,f}} , \qquad
 \operatorname{ch} \widetilde{\mathbf{M}}_i = \sum_{f = 1}^{n^\text{af}_i} {\rm e}^{\widetilde{m}_{i,f}} .
\]

\subsection*{Contour integral formula}

The contour integral form of the instanton partition function is given by applying the index functor to the instanton part of the total contribution
\begin{gather}
 Z_\gamma
 = \mathbb{I}\bigg[ \sum_{i\in\Gamma_0} \big( \mathbf{V}_i + \mathbf{H}_i + \widetilde{\mathbf{H}}_i \big) + \sum_{e \in \Gamma_1} \mathbf{H}_e \bigg]_\text{inst}
 = \frac{1}{\underline{k}!} \oint_{\mathbb{T}_\mathbf{K}} \dd{\underline{\phi}} \prod_{i \in \Gamma_0} z_{i}^\text{vec} z_{i}^\text{f} z_{i}^\text{af} \prod_{e \in \Gamma_1} z_e^\text{bf},
\label{eq:LMNS_quiver}
\end{gather}
where we denote the Cartan torus of ${\rm GL}(\mathbf{K}$) by $\mathbb{T}_\mathbf{K}$ and
\[
 \underline{k}! = \prod_{i \in \Gamma_0} k_i! , \qquad
 \dd\underline{\phi} = \prod_{i \in \Gamma_0} \prod_{I = 1}^{k_i} \frac{\dd \phi_{i,I}}{2 \pi \iota} .
\]
Each factor in the integrand is given by
\begin{gather*}
 z_i^\text{vec} = \frac{[-\epsilon_{12}]^{k_i}}{[-\epsilon_{1,2}]^{k_i}} \prod_{I = 1}^{k_i} \frac{1}{P_i(\phi_{i,I}) \widetilde{P}_i(\phi_{i,I} + \epsilon_{12})} \prod_{I < J}^{k_i} \mathscr{S}(\phi_{i,IJ})^{-1} \mathscr{S}(\phi_{i,IJ}+\epsilon_{12})^{-1} 
 , \\
 z_{e:i \to j}^\text{bf} = \prod_{I = 1}^{k_j} P_i(\phi_{j,I} + m_e) \prod_{I = 1}^{k_i} \widetilde{P}_j(\phi_{i,I} - m_e + \epsilon_{12}) \prod_{\substack{I = 1,\ldots, k_j \\ J = 1,\ldots,k_i}} \mathscr{S}(\phi_{j,I} - \phi_{i,J} + m_e)
 , \\
 z_i^\text{f} = \prod_{I = 1}^{k_i} P_{i}^\text{f}(\phi_{i,I})
 , \qquad
 z_i^\text{af} = \prod_{I = 1}^{k_i} \widetilde{P}_{i}^\text{f}(\phi_{i,I} + \epsilon_{12}),
\end{gather*}
where we denote $\phi_{i,IJ} = \phi_{i,I} - \phi_{i,J}$, and the gauge and matter polynomials are given by the corresponding characteristic polynomials
\begin{alignat*}{3}
 &P_i(x) = P_{\mathbf{N}_i}(x) = \prod_{\alpha = 1}^{n_i} [x - a_{i,\alpha}] , \qquad&&
 \widetilde{P}_i(x) = \widetilde{P}_{\mathbf{N}_i}(x) = \prod_{\alpha = 1}^{n_i} [- x + a_{i,\alpha}] , &\\
 &P_i^\text{f}(x) = {P}_{\mathbf{M}_i}(x) = \prod_{f = 1}^{n_i^\text{f}} [x - m_{i,f}] , \qquad&&
 \widetilde{P}_i^\text{f}(x) = \widetilde{P}_{\widetilde{\mathbf{M}}_i}(x) = \prod_{f = 1}^{n_i^\text{af}} [- x + \widetilde{m}_{i,f}] .&
\end{alignat*}
In addition, we can further incorporate the Chern--Simons term for 5d theory
\[
 z_i^\text{CS} = \prod_{I=1}^{k_i} {\rm e}^{\kappa_i \phi_{i,I}} , \qquad
 i \in \Gamma_0 ,
\]
where we denote the Chern--Simons level by $(\kappa_i)_{i \in \Gamma_0}$.
The total partition function is given by summing over the topological sectors.
In this case, we have the counting parameter for each node $\underline{\mathfrak{q}} = (\mathfrak{q}_i)_{i \in \Gamma_0}$, which yields the total partition function
\begin{gather}
 Z = \sum_{\underline{k}} \underline{\mathfrak{q}}^{\underline{k}} Z_{\gamma} , \qquad
 \underline{\mathfrak{q}}^{\underline{k}} = \prod_{i \in \Gamma_0} \mathfrak{q}_i^{k_i} . \label{eq:tot_pf_quiv}
\end{gather}

\subsection*{Localization formula}

We can similarly apply the equivariant localization formula to quiver gauge theory in general.
In this case, the fixed points in the moduli space $\mathfrak{M}_\gamma$ are again labeled by a set of partitions
\[
 \lambda = (\lambda_{i,\alpha})_{i \in \Gamma_0, \alpha = 1,\ldots,n_i} , \qquad
 |\lambda_i| = \sum_{\alpha = 1}^{n_i} |\lambda_{i,\alpha}| = k_i .
\]
Then, the observable sheaf is given through the partial reduction of the universal sheaf by
\[
 \mathbf{Y}_i = \mathbf{P}_1 \mathbf{X}_i,
\]
where the corresponding character is given by
\[
 \operatorname{ch} \mathbf{X}_i = \sum_{x \in \mathcal{X}_i} x , \qquad
 \mathcal{X}_i = \big\{ {\rm e}^{a_{i,\alpha}} q_1^{k-1} q_2^{\lambda_{i,\alpha,k}} \big\}_{i \in \Gamma_0,\alpha = 1,\ldots,n_i,k=1,\ldots,\infty} .
\]
We also define the empty configuration
\[
 \mathring{\mathcal{X}}_i = \big\{ {\rm e}^{a_{i,\alpha}} q_1^{k-1} \big\}_{i \in \Gamma_0,\alpha = 1,\ldots,n_i,k=1,\ldots,\infty} .
\]
We denote the fixed point configuration by $\mathcal{X} = (\mathcal{X}_i)_{i \in \Gamma_0} \in \mathfrak{M}^\mathsf{T} = \bigsqcup_{\underline{k}} \mathfrak{M}_{\underline{n},\underline{k}}^\mathsf{T}$.
The vector multiplet and the hypermultiplet contributions are written as follows:
\begin{alignat*}{3}
& \mathbf{V}_i = \frac{\mathbf{P}_1^\vee}{\mathbf{P}_2} \mathbf{X}_i^\vee \mathbf{X}_i , \qquad &&
 \mathbf{H}_{e\colon i \to j} = - \mathbf{M}_e \frac{\mathbf{P}_1^\vee}{\mathbf{P}_2} \mathbf{X}_i^\vee \mathbf{X}_j , &\\
 &\mathbf{H}_i = - \frac{\mathbf{M}_i^\vee \mathbf{X}_i}{\mathbf{P}_2}
 , \qquad &&
 \widetilde{\mathbf{H}}_i = - \mathbf{Q}_{12}^\vee \frac{\mathbf{X}_i^\vee \widetilde{\mathbf{M}}_i}{\mathbf{P}_2} .&
\end{alignat*}
In particular, the vector multiplet and the bifundamental contributions are combined into the following form:
\[
 \sum_{i \in \Gamma_0} \mathbf{V}_i + \sum_{e \in \Gamma_1} \mathbf{H}_e
 = \frac{\mathbf{P}_1^\vee}{\mathbf{P}_2} \sum_{i,j \in \Gamma_0} \mathbf{X}_i^\vee \mathbf{c}^{+\vee}_{ij} \mathbf{X}_j
 = \frac{\mathbf{P}_1^\vee}{\mathbf{P}_2} \mathbf{X}^\vee \mathbf{c}^{+\vee} \mathbf{X} ,
\]
where $\mathbf{c}^+$ is the half $q$-Cartan matrix~\eqref{eq:half_Cartan_mat}.
Thus, the fixed point contribution of the partition function is given by
\[
 Z_{\mathcal{X}} = \mathbb{I}\bigg[ \sum_{i\in\Gamma_0} \big( \mathbf{V}_i + \mathbf{H}_i + \widetilde{\mathbf{H}}_i \big) + \sum_{e \in \Gamma_1} \mathbf{H}_e \bigg]_\mathcal{X},
\]
and the total partition function is given by summation over all the fixed point contributions
\[
 Z = \sum_{\mathcal{X} \in \mathfrak{M}^\mathsf{T}} \underline{\mathfrak{q}}^{\underline{k}(\mathcal{X})} Z_{\mathcal{X}}.
\]
The topological number is given by $\underline{\mathfrak{q}}^{\underline{k}(\mathcal{X})} = \prod_{i \in \Gamma_0} \mathfrak{q}_i^{k_i}$ for $\mathcal{X} \in \mathfrak{M}_{\underline{n},\underline{k}}^\mathsf{T}$.

\subsection*{Including a defect}

As discussed in Section~\ref{sec:defect}, we may similarly incorporate a defect operator to quiver gauge theory.
In this case, we define the corresponding $\mathscr{Y}$-functions
\[
 \mathscr{Y}_i(x) = \mathbb{I}\big[\mathbf{Y}_i^\vee \mathbf{x}\big] , \qquad
 \widetilde{\mathscr{Y}}_i(x) = \mathbb{I}\big[\mathbf{x}^\vee \mathbf{Y}_i\big] .
\]
Denoting another framing space and the character
\[
 \mathbf{N}_i' = \mathbb{C}^{n_i'} , \qquad
 \mathbf{N}' = (\mathbf{N}_i')_{i \in \Gamma_0} , \qquad
 \operatorname{ch} \mathbf{N}_i' = \sum_{\beta = 1}^{n_i'} {\rm e}^{b_{i,\beta}} ,
\]
we have
\[
 \mathscr{Y}_{\mathbf{N}'} = \mathbb{I}\bigg[ \sum_{i \in \Gamma_0} \mathbf{Y}_i^\vee \mathbf{N}_i'\bigg] = \prod_{i \in \Gamma_0} \prod_{\beta = 1}^{n_i'} \mathscr{Y}(b_{i,\beta}) .
\]
Hence, for the topological data $\gamma = (\underline{n},\underline{n}',\underline{k}) = (n_i,n_i',k_i)_{i \in \Gamma_0}$, the total partition function is given as follows:
\begin{align*}
 Z_\gamma = \mathbb{I}\bigg[ \sum_{i\in\Gamma_0} \big( \mathbf{V}_i + \mathbf{H}_i + \widetilde{\mathbf{H}}_i + \mathbf{Y}_i^\vee \mathbf{N}_i' \big) + \sum_{e \in \Gamma_1} \mathbf{H}_e \bigg]_\text{inst}
 = \frac{1}{\underline{k}!} \oint_{\mathbb{T}_\mathbf{K}} \dd{\underline{\phi}} \mathscr{Y}_{\mathbf{N}'} \prod_{i \in \Gamma_0} z_{i}^\text{vec} z_{i}^\text{f} z_{i}^\text{af} \prod_{e \in \Gamma_1} z_e^\text{bf} .
\end{align*}
The instanton sum gives rise to the (average of) $qq$-character, which is a pole-free function of the dual Coulomb parameters $b_i = \sum_{\beta = 1}^{n_i'} b_{i,\beta}$ for $i \in \Gamma_0$.

\subsubsection[A\_1 quiver: SQCD]{$\boldsymbol{A_1}$ quiver: SQCD}

We start with a minimal example with the hypermultiplet, which is 4d $\mathcal{N}=2$ (5d $\mathcal{N}=1$) SQCD with $n^\text{f}$ fundamental and $n^\text{af}$ antifundamental matters.
We need to have the condition $n^\text{f} + n^\text{af} = 2n$ to realize the corresponding 6d $\mathcal{N}=(1,0)$ theory.
Applying the formula~\eqref{eq:LMNS_quiver}, we obtain the contour integral form of the $k$-instanton partition function of U($n$) SQCD as follows:
\[
 Z_\gamma = \frac{1}{k!} \frac{[-\epsilon_{12}]^k}{[-\epsilon_{1,2}]^{k}} \oint_{\mathbb{T}_\mathbf{K}} \prod_{I=1}^k \frac{\dd{\phi}_I}{2 \pi \iota} \frac{P^\text{f}(\phi_I) \widetilde{P}^\text{f}(\phi_I+\epsilon_{12})}{P(\phi_I) \widetilde{P}(\phi_I + \epsilon_{12})} \prod_{I \neq J}^k \mathscr{S} (\phi_{IJ})^{-1} ,
\]
where we denote the gauge polynomial by $P(x) = P_{\mathbf{N}}(x)$ and its dual $\widetilde{P}(x) = \widetilde{P}_{\mathbf{N}}(x)$.
In this case, in order to have a converging integral (no pole at $\phi \to \infty$), there exists a restriction on the matter content given by $n^\text{f} + n^\text{af} \le 2 n$.

\subsubsection[widehat A\_0 quiver]{$\boldsymbol{\widehat{A}_0}$ quiver}

We consider quiver gauge theory associated with $\widehat{A}_0$ quiver, which is known as 4d $\mathcal{N}=2^*$ (5d $\mathcal{N}=1^*$/6d $\mathcal{N}=(1,0)^*$) theory.
The matter contents are the vector multiplet and the hypermultiplet in the adjoint representation of the gauge group.
Denoting the adjoint mass parameter by $m$, and applying the general formula~\eqref{eq:LMNS_quiver}, the $k$-instanton partition function of~U($n$) gauge theory with the adjoint matter is given by
\[
 Z_\gamma = \frac{1}{k!} \left( \frac{[-\epsilon_{12}]}{[-\epsilon_{1,2}]} \mathscr{S}(m) \right)^k
 \oint_{\mathbb{T}_\mathbf{K}} \prod_{I = 1}^k \frac{\dd{\phi}_I}{2 \pi \iota} \frac{P(\phi_I + m) \widetilde{P}(\phi_I - m + \epsilon_{12})}{P(\phi_I) \widetilde{P}(\phi_I + \epsilon_{12})} \prod_{I \neq J}^k \frac{\mathscr{S}(\phi_{IJ} + m)}{\mathscr{S} (\phi_{IJ})} .
\]
We remark that, for the 5d/6d convention, the gauge polynomial part is written as follows:
\begin{align*}
 \frac{P(\phi + m) \widetilde{P}(\phi - m + \epsilon_{12})}{P(\phi) \widetilde{P}(\phi + \epsilon_{12})}
 & = {\rm e}^{- n m} \frac{P(\phi + m) {P}(\phi - m + \epsilon_{12})}{P(\phi) {P}(\phi + \epsilon_{12})} \nonumber \\
 & = {\rm e}^{- n m} \prod_{\alpha = 1}^n \frac{[x - a_\alpha + m][x - a_\alpha - m + \epsilon_{12}]}{[x - a_\alpha][x - a_\alpha + \epsilon_{12}]} .
\end{align*}
In the 4d convention, there is no prefactor.

We use the notation
\[
 (\epsilon_3, \epsilon_{4}) = (-m, m-\epsilon_{12}) , \qquad
 (q_3, q_4) = ({\rm e}^{\epsilon_3}, {\rm e}^{\epsilon_4}) ,
\]
for which we have the condition
\begin{gather}
 \sum_{i=1}^4 \epsilon_i = 0 , \qquad
 \prod_{i=1}^4 q_i = 1 .
 \label{eq:CY4_cond}
\end{gather}
As mentioned below, this theory has a geometric origin in $\mathbb{C}^4$, and in this context, this condition is called the Calabi--Yau condition:
The associated $\Omega$-background parameters take a value in the Cartan subalgebra of $\mathfrak{u}(4)$, which is reduced to $\mathfrak{su}(4)$ under the condition~\eqref{eq:CY4_cond}.

Similarly, the measure part is rewritten as follows:
\begin{gather*}
 \prod_{I \neq J}^k \frac{\mathscr{S}(\phi_{IJ} + m)}{\mathscr{S} (\phi_{IJ})}
 = \prod_{I < J}^k \frac{\mathscr{S}(\phi_{IJ} + m)}{\mathscr{S} (\phi_{IJ})} \frac{\mathscr{S}(\phi_{IJ} - m + \epsilon_{12})}{\mathscr{S} (\phi_{IJ} + \epsilon_{12})}
 = \prod_{I < J}^k \frac{[\phi_{IJ} \pm \epsilon_{12,23,31}]}{[\phi_{IJ} \pm \epsilon_{1,2,3,4}]} [\phi_{IJ}]^2 .
\end{gather*}
Hence, the $k$-instanton partition function is written as follows:
\begin{gather}
 Z_\gamma = \frac{q_3^{nk}}{k!} \frac{[-\epsilon_{12,23,31}]^k}{[-\epsilon_{1,2,3}]^{k}[-\epsilon_{123}]^k}
 \oint_{\mathbb{T}_\mathbf{K}} \prod_{I = 1}^k \frac{\dd{\phi}_I}{2 \pi \iota} \prod_{\alpha = 1}^n \mathscr{S}_{34}(\phi_I - a_\alpha) \prod_{I < J}^k \frac{[\phi_{IJ} \pm \epsilon_{12,23,31}]}{[\phi_{IJ} \pm \epsilon_{1,2,3,4}]} [\phi_{IJ}]^2 ,
 \label{eq:epsilon34}
\end{gather}
which implies a natural realization of $\widehat{A}_0$ theory in $\mathbb{C}^4$.
See Section~\ref{sec:geometry} for details.
This integral computes the $\chi_{q_3}$-genus and the elliptic genus of the instanton moduli space $\mathfrak{M}_{\gamma}$ for the~5d and~6d conventions, respectively.

Moreover, we may insert a codimension-four defect to this case.
Inserting the $\mathscr{Y}$-functions, we obtain the following integral formula~\cite{Kimura:2020jxl}:
\begin{gather*}
 Z_\gamma = \frac{q_3^{nk}}{k!} \frac{[-\epsilon_{12,23,31}]^k}{[-\epsilon_{1,2,3}]^{k}[-\epsilon_{123}]^k}
 \oint_{\mathbb{T}_\mathbf{K}} \dd{\underline{\phi}} \mathscr{Y}_{\mathbf{N}'} \prod_{\substack{\alpha = 1,\ldots,n \\ I = 1,\ldots,k}} \mathscr{S}_{34}(\phi_I - a_\alpha) \prod_{I < J}^k \frac{[\phi_{IJ} \pm \epsilon_{12,23,31}]}{[\phi_{IJ} \pm \epsilon_{1,2,3,4}]} [\phi_{IJ}]^2
\\
\hphantom{Z_\gamma}{} =
 \frac{q_3^{nk}}{k!} \frac{[-\epsilon_{12,23,31}]^k}{[-\epsilon_{1,2,3}]^{k}[-\epsilon_{123}]^k}
 \oint_{\mathbb{T}_\mathbf{K}} \dd{\underline{\phi}} \prod_{\substack{\alpha = 1,\ldots,n \\ \beta = 1,\ldots,n'}} [b_\beta - a_\alpha] \prod_{\substack{\alpha = 1,\ldots,n \\ I = 1,\ldots,k}} \mathscr{S}_{34}(\phi_I - a_\alpha) \\
\hphantom{Z_\gamma=}{} \times
 \prod_{\substack{\beta = 1,\ldots,n' \\ I = 1,\ldots,k}} \mathscr{S}_{12}(\phi_I - b_\beta)
 \prod_{I < J}^k \frac{[\phi_{IJ} \pm \epsilon_{12,23,31}]}{[\phi_{IJ} \pm \epsilon_{1,2,3,4}]} [\phi_{IJ}]^2 ,
\end{gather*}
which is symmetric under $(a_\alpha,b_\beta)$ and $(\epsilon_{1,2},\epsilon_{3,4})$.

\subsubsection[widehat A\_\{r-1\} quiver]{$\boldsymbol{\widehat{A}_{r-1}}$ quiver}\label{sec:Ar^_quiver}

The next example is quiver gauge theory of type $\widehat{A}_{r-1}$, which is a cyclic quiver gauge theory of $r$ gauge nodes.
Although we have $r$ possibly different bifundamental mass parameters in general, we apply a gauge choice such that all the mass parameters are identical, $m_{i \to i+1} = m$ for $i = 0,\ldots,r-1$.
The node index is now understood as $i \equiv i + r$, hence $i \in \mathbb{Z}/r\mathbb{Z}$.
We also denote $\mathbb{Z}/r\mathbb{Z} = \mathbb{Z}_r$.

Applying the index functor, we have the contour integral form of the instanton partition function
\begin{gather*}
\begin{split}
& Z_\gamma = \frac{1}{\underline{k}!} \frac{[-\epsilon_{12}]^{|k|}}{[-\epsilon_{1,2}]^{|k|}}
 \oint_{\mathbb{T}_\mathbf{K}} \dd{\underline{\phi}} \prod_{i=0}^{r-1} \Bigg[ \prod_{I = 1}^{k_i} \frac{P_{i-1}(\phi_{i,I} + m) \widetilde{P}_{i+1}(\phi_{i,I} - m + \epsilon_{12})}{P_i(\phi_{i,I}) \widetilde{P}_i(\phi_{i,I} + \epsilon_{12})}
 \nonumber \\
& \hphantom{Z_\gamma =}{}
 \times
 \frac{ \prod_{\substack{I=1,\ldots,k_{i+1} \\ J = 1,\ldots,k_{i}}} \mathscr{S}(\phi_{i+1,I} - \phi_{i,J} + m)}{ \prod_{I \neq J}^{k_i} \mathscr{S} (\phi_{i,I} - \phi_{i,J}) } \Bigg] ,
\end{split}
\end{gather*}
where we denote $|k| = \sum_{i=0}^{r-1} k_i$.
The gauge polynomial part is written for the 5d/6d convention as follows:
\begin{gather*}
 \prod_{i=0}^{r-1} \frac{P_{i-1}(\phi + m) \widetilde{P}_{i+1}(\phi - m + \epsilon_{12})}{P_i(\phi) \widetilde{P}_i(\phi + \epsilon_{12})}
 \nonumber \\
\qquad = \prod_{i=0}^{r-1} (-q_{12})^{ (n_{i-1} - n_i)k_i} q_3^{ n_{i-1} k_i} \phi^{n_{i-1} - n_i} {\rm e}^{(k_{i+1} - k_i) a_i } \frac{P_{i-1}(\phi + m) {P}_{i+1}(\phi - m + \epsilon_{12})}{P_i(\phi) {P}_i(\phi + \epsilon_{12})} ,
\end{gather*}
where we denote $a_i = \sum_{\alpha = 1}^{n_i} a_{i,\alpha}$.
We remark that the bifundamental factor is written as
\[
 \mathscr{S}(\phi_{i+1,I} - \phi_{i,J} + m) = \frac{[\phi_{i+1,I} - \phi_{i,J} - \epsilon_{13}][\phi_{i+1,I} - \phi_{i,J} + \epsilon_{14}]}{[\phi_{i+1,I} - \phi_{i,J} - \epsilon_3][\phi_{i+1,I} - \phi_{i,J} + \epsilon_4]}.
\]
Hence, imposing the balancing (conformal) condition and the special unitary condition, $n_i = n$, $a_i = 0$ for $i = 0,\ldots,r-1$, we obtain
\begin{gather}
 Z_\gamma = \frac{q_3^{n |k|}}{\underline{k}!} \frac{[-\epsilon_{12}]^{|k|}}{[-\epsilon_{1,2}]^{|k|}}
 \oint_{\mathbb{T}_\mathbf{K}} \dd{\underline{\phi}} \prod_{i=0}^{r-1} \Bigg[ \prod_{I = 1}^{k_i} \frac{P_{i-1}(\phi_{i,I} - \epsilon_3) {P}_{i+1}(\phi_{i,I} - \epsilon_4)}{P_i(\phi_{i,I}) {P}_i(\phi_{i,I} - \epsilon_{34})}
 \nonumber \\
 \hphantom{Z_\gamma = \frac{q_3^{n |k|}}{\underline{k}!} \frac{[-\epsilon_{12}]^{|k|}}{[-\epsilon_{1,2}]^{|k|}}
 \oint_{\mathbb{T}_\mathbf{K}}}{}
 \times
 \frac{\prod_{\substack{I=1,\ldots,k_{i+1} \\ J = 1,\ldots,k_{i}}} \mathscr{S}(\phi_{i+1,I} - \phi_{i,J} - \epsilon_3)}{ \prod_{I < J}^{k_i} \mathscr{S} (\phi_{i,IJ} ) \mathscr{S} (\phi_{i,IJ} - \epsilon_{34}) } \Bigg] . \label{eq:Ar-1^_quiver}
\end{gather}
The total partition function is given by summing over all the topological sectors as in \eqref{eq:tot_pf_quiv}.

\subsection{Orbifold instanton partition function}\label{sec:orb_inst}

We extend the formulation of the instanton partition function to the orbifold case, $\mathcal{S} = \mathbb{C}^2/\Upsilon$, where $\Upsilon$ is a finite subgroup of SU(2).
For the moment, we focus on the A-type theory, $\Upsilon = \mathbb{Z}_p = \mathbb{Z}/p\mathbb{Z}$.
We also denote $\widehat{A}_{p-1}$ quiver by $\Upsilon$ under the McKay--Nakajima correspondence, hence $\Upsilon_0 = \{0,\ldots,p-1\}$.

In order to discuss the instanton partition function of $\Gamma$-quiver gauge theory on the orbifold, we first decompose the vector spaces as follows:
\[
 \mathbf{N}_i = \bigoplus_{j \in \Upsilon_0} \mathbf{N}_i^j \otimes \mathcal{R}_j , \qquad
 \mathbf{K}_i = \bigoplus_{j \in \Upsilon_0} \mathbf{K}_i^j \otimes \mathcal{R}_j , \qquad i \in \Gamma_0 ,
\]
where we denote the one-dimensional irreducible representation of $\Upsilon = \mathbb{Z}_p$ by $\mathcal{R}_j$ for $j = 0,\ldots,p-1$.
In particular, we denote the trivial representation by $\mathcal{R}_0$.
Each vector space is given by
\[
 \mathbf{N}_i^j = \mathbb{C}^{n_i^j} , \qquad \mathbf{K}_i^j = \mathbb{C}^{k_i^j} ,
\]
obeying $\sum_{j \in \Upsilon_0} n_i^j = n_i$, $\sum_{j \in \Upsilon_0} k_i^j = k_i$, and we write by $\gamma = (\underline{n},\underline{k}) = \big(n_i^j,k_i^j\big)_{i \in \Gamma_0,j = \Upsilon_0}$ the topological data.
We denote the characters of these vector spaces by
\[
 \operatorname{ch} \mathbf{N}_i^j = \sum_{\alpha = 1}^{n_i^j} {\rm e}^{a_{i,\alpha}^j} , \qquad
 \operatorname{ch} \mathbf{K}_i^j = \sum_{I = 1}^{k_i^j} {\rm e}^{\phi_{i,I}^j} .
\]
Defining the coloring index $c_i\colon \{1,\ldots,n_i\} \to \{ 0,\ldots,p-1 \}$ for each $i \in \Gamma_0$, we may identify $\big\{ a_{i,\alpha}^j, \alpha = 1,\ldots,n_i^j \big\} = \{ a_{i,\alpha}, \alpha = 1,\ldots,n_i \mid c_i(\alpha) = j \}$ where $n_i^j = \#\{ \alpha = 1,\ldots,n_i \mid c_i(\alpha) = j \}$.

In addition, we consider the following $\mathbb{Z}_p$ action for the spacetime coordinates:
\[
 \mathbb{Z}_p \colon \
 \begin{pmatrix}
 z_1 \\ z_2
 \end{pmatrix}
 \longrightarrow
 \begin{pmatrix}
 {\rm e}^{2 \pi \iota (p-1) / p} & 0 \\ 0 & {\rm e}^{2 \pi \iota / p}
 \end{pmatrix}
 \begin{pmatrix}
 z_1 \\ z_2
 \end{pmatrix} , 
\]
which correspond to the assignment of the irreducible representation
\begin{gather}
\mathbf{Q} = (\mathbf{Q}_1 \otimes \mathcal{R}_{p-1}) \oplus ( \mathbf{Q}_2 \otimes \mathcal{R}_1 ).
\label{eq:Q_rep}
\end{gather}
We remark that $\mathcal{R}_{p-1} = {\mathcal{R}}_1^\vee$, and the two-dimensional representation that can be embedded with that for SU(2) is given by ${\mathcal{R}}_1^\vee \oplus {\mathcal{R}}_1$.
Then, the instanton partition function is obtained from the $\Upsilon$-invariant sector of the original contributions.
For simplicity, we consider $\Gamma=A_1$ quiver for the moment.
Recalling that
\[
 (\mathcal{R}_i \otimes \mathcal{R}_j)_\Upsilon = \delta_{i+j,0} ,
\]
the vector multiplet contribution is given as follows:
\begin{align}
 \mathbf{V}_{\text{inst}}^\Upsilon & = \bigl( - \mathbf{N}^\vee \mathbf{K} - \mathbf{Q}_{12}^\vee \mathbf{K}^\vee \mathbf{N} + \mathbf{P}_{12}^\vee \mathbf{K}^\vee \mathbf{K} \bigr)_\Upsilon \nonumber \\
 & = \sum_{j \in \Upsilon_0} \bigl( - \mathbf{N}^{j\vee} \mathbf{K}^j - \mathbf{Q}_{12}^\vee \mathbf{K}^{j\vee} \mathbf{N}^j + \big( 1 + \mathbf{Q}_{12}^\vee \big) \mathbf{K}^{j\vee} \mathbf{K}^j - \mathbf{Q}_1^\vee \mathbf{K}^{j\vee} \mathbf{K}^{j-1} - \mathbf{Q}_2^\vee \mathbf{K}^{j\vee} \mathbf{K}^{j+1} \bigr) \nonumber \\
 & =
 - \mathbf{N}^\vee \mathbf{K} - \mathbf{Q}_{12}^\vee \mathbf{K}^\vee \mathbf{N} + \mathbf{K}^\vee \mathbf{c}^\dag \mathbf{K} , \label{eq:vect_contribution_upsilon}
\end{align}
where we apply the vector notation in the last line with the $q$-deformed Cartan matrix of type $\widehat{A}_{p-1}$ associated with $\Upsilon = \mathbb{Z}_p$,\footnote{The $q$-Cartan matrix used in this section is with respect to the orbifold structure $\Upsilon$ not to be confused with that for quiver $\Gamma$~\eqref{eq:half_Cartan_mat}.}
\begin{gather}
 \mathbf{c}_{ij} = (1 + \mathbf{Q}_{12}) \delta_{i,j} - \mathbf{Q}_1 \delta_{i,j+1} - \mathbf{Q}_2 \delta_{i+1,j} , \qquad
 \mathbf{c}^\dag := \mathbf{c}^{\vee\text{T}} = \mathbf{Q}_{12}^\vee \mathbf{c} . \label{eq:qCartan_Ap-1}
\end{gather}
Now the spacetime equivariant parameters $\epsilon_{1,2}$ play a role of the bifundamental mass parameters.

In general, assigning the representation $\mathcal{R}_k$ to $\mathbf{Q}_2$, we have $\mathbf{Q}_2 \ \to \ \mathbf{Q}_2 \delta_{i+k,j}$ in the $q$-Cartan matrix.
Namely, the irreducible representation $\mathcal{R}_k$ corresponds to the matrix $\mathscr{R}_k$ given by
\begin{gather}
 \mathcal{R}_k \longrightarrow \mathscr{R}_k =
 \begin{pmatrix}
 &&&\overbrace{1}^{p-k}&&&& \\
 &&&&1&&& \\
 &&&&&\ddots&& \\
 &&&&&&\ddots& \\
 &&&&&&&1 \\
 1&&&&&&& \\
 &\ddots&&&&&& \\
 &&1&&&&&
 \end{pmatrix} . \label{eq:Rk_matrix}
\end{gather}
Apparently, we have $(\mathscr{R}_k)^p = \mathbbm{1}_p$.
We remark that the inverse matrix is given by $\mathscr{R}_k^{-1} = \mathscr{R}_k^\text{T} = \mathscr{R}_{p-k}$.
In this notation, we may write the $q$-Cartan matrix \eqref{eq:qCartan_Ap-1} by
\[
 \mathbf{c} = (1 + \mathbf{Q}_{12}) \otimes \mathbbm{1} - \mathbf{Q}_1 \otimes \mathscr{R}_{p-1} - \mathbf{Q}_2 \otimes \mathscr{R}_{1} = (1 - \mathbf{Q}_1 \otimes \mathscr{R}_{p-1}) (1 - \mathbf{Q}_2 \otimes \mathscr{R}_{1}) .
\]
See Appendix~\ref{sec:Cartan_gen} for the $q$-Cartan matrix with a generic assignment of the irreducible representation.

We also use the notation $\mathbf{c}_\Upsilon$ to specify the quiver dependence of the $q$-Cartan matrix.
We denote the character of the $q$-Cartan matrix by $c$.
The determinant of the $q$-Cartan matrix of type $\widehat{A}_{p-1}$ is given by
\[
 \det \mathbf{c} = \mathbf{P}_{12}^{[p]} , \qquad \det c = \big(1 - q_1^p\big)\big(1 - q_2^p\big) .
\]
Thus, the $q$-Cartan matrix ${c}$ is invertible if $q_{1,2} \neq 1$ ($\epsilon_{1,2} \neq 0$).%
\footnote{Precisely speaking, we should impose $q_{1,2} \neq \exp(2 r \pi \iota / p)$ for $r = 0,\ldots,p-1$.
The role of the root of unity limit in the context of gauge theory is discussed in~\cite{Kimura:2011zf,Kimura:2011gq}.}
We denote the inverse of the $q$-Cartan matrix by $\tilde{\mathbf{c}}$.
In the case $\Upsilon = \widehat{A}_0$, it becomes $\mathbf{c} = \mathbf{P}_{12}$, and the inverse is simply given by $\tilde{\mathbf{c}} = 1/\mathbf{P}_{12}$.
The elements of the inverse matrix $\tilde{{c}}$ of type $\widehat{A}_{p-1}$ is then given as follows:
\begin{align*}
 \tilde{c}_{ij} & = \sum_{\substack{x,y \in \Upsilon_0 \\ x-y \equiv j-i~(\text{mod}~p)}} \frac{q_1^x q_2^y}{\big(1-q_1^p\big)\big(1-q_2^p\big)} \nonumber \\
 & =
 \begin{cases}
 \displaystyle
 \frac{q_2^s \big(1 + q_{12} + \cdots + q_{12}^{p-s-1}\big) + q_1^{p-s} \big(1 + q_{12} + \cdots + q_{12}^{s-1}\big)}{\big(1-q_1^p\big)\big(1-q_2^p\big)}, & s:=i-j \ge 0, \vspace{1mm}\\
 \displaystyle
 \frac{ q_2^{p-t} \big(1 + q_{12} + \cdots + q_{12}^{t-1}\big) + q_1^t \big(1 + q_{12} + \cdots + q_{12}^{p-t-1}\big)}{\big(1-q_1^p\big)\big(1-q_2^p\big)}, & t:=j-i \ge 0,
 \end{cases}
\end{align*}
where we denote $\Upsilon_0 = \{ 0,\ldots,p-1 \}$.

We also decompose the observable sheaf with respect to the irreducible representation of $\Upsilon$,
\[
 \mathbf{Y}_i = \bigoplus_{j \in \Upsilon_0} \mathbf{Y}_i^j \otimes \mathcal{R}_j , \qquad
 i \in \Gamma_0 .
\]
We denote the set of observable sheaves by $\mathbf{Y} = \big(\mathbf{Y}_i^j\big)_{i \in \Gamma_0, j \in \Upsilon_0}$.
Then, we have the following expression,
\[
 \mathbf{Y}_i^j = \mathbf{N}_i^j - \sum_{j' \in \Upsilon_0} \mathbf{c}_{jj'} \mathbf{K}_i^{j'} ,
\]
for which we may write $\mathbf{Y}_i = \mathbf{N}_i - \mathbf{c} \mathbf{K}_i$ in the vector notation.
The vector multiplet contribution is thus given by
\begin{gather}
 \mathbf{V}_i^\Upsilon = \mathbf{Y}_i^\vee \tilde{\mathbf{c}} \mathbf{Y}_i = \sum_{j,j' \in \Upsilon_0} \mathbf{Y}_i^{j\vee} \tilde{\mathbf{c}}_{jj'} \mathbf{Y}_i^{j'} ,
 \label{eq:vect_contribution_upsilon_full}
\end{gather}
which reproduces the previous result~\eqref{eq:vect_contribution_upsilon} for the instanton part.

In order to construct quiver gauge theory on orbifolds, we consider the hypermultiplet contribution similarly to Section~\ref{sec:quiver_gauge_theory}.
Using the $q$-Cartan matrix, we have
\begin{subequations}\label{eq:hyp_contribution_upsilon_full}
\begin{gather}
 \mathbf{H}_{e\colon i \to i'}^\Upsilon = - \mathbf{M}_e \mathbf{Y}_i^\vee \tilde{\mathbf{c}} \mathbf{Y}_{i'}
 = - \mathbf{M}_e \big( \mathbf{N}_i^\vee \tilde{\mathbf{c}} \mathbf{N}_{i'}
 - \mathbf{N}_i^\vee \mathbf{K}_{i'} - \mathbf{Q}_{12}^\vee \mathbf{K}_i^\vee \mathbf{N}_{i'} + \mathbf{K}_i^\vee \mathbf{c}^\dag \mathbf{K}_{i'} \big) , \\
 \mathbf{H}_{i}^\Upsilon = - \mathbf{M}_i^\vee \tilde{\mathbf{c}} \mathbf{Y}_i = - \mathbf{M}_i^\vee \tilde{\mathbf{c}} \mathbf{N}_i + \mathbf{M}_i^\vee \mathbf{K}_i , \\
 \widetilde{\mathbf{H}}_{i}^\Upsilon = - \mathbf{Y}_i^\vee \tilde{\mathbf{c}} \widetilde{\mathbf{M}}_i = - \mathbf{N}_i^\vee \tilde{\mathbf{c}} \widetilde{\mathbf{M}}_i + \mathbf{Q}_{12}^\vee \mathbf{K}_i^\vee \widetilde{\mathbf{M}}_i .
\end{gather}
\end{subequations}
Applying the index functor to the instanton part as in \eqref{eq:LMNS_quiver}, we obtain the contour integral formula of the orbifold instanton partition function for the topological data $\gamma = \big(n_i^j,k_i^j\big)_{i \in \Gamma_0, j \in \Upsilon_0}$,
\begin{gather*}
 Z_\gamma
 = \mathbb{I}\bigg[ \sum_{i\in\Gamma_0} \big( \mathbf{V}_i + \mathbf{H}_i + \widetilde{\mathbf{H}}_i \big) + \sum_{e \in \Gamma_1} \mathbf{H}_e \bigg]_\text{inst}
 = \frac{1}{\underline{k}!} \oint_{\mathbb{T}_\mathbf{K}} \dd{\underline{\phi}} \prod_{i \in \Gamma_0} z_{i}^\text{vec} z_{i}^\text{f} z_{i}^\text{af} \prod_{e \in \Gamma_1} z_e^\text{bf},
\end{gather*}
where we denote
\[
 \underline{k}! = \prod_{\substack{i \in \Gamma_0 \\ j \in \Upsilon_0}} k_i^j! , \qquad
 \dd\underline{\phi} = \prod_{\substack{i \in \Gamma_0 \\ j \in \Upsilon_0}} \prod_{I = 1}^{k_i^j} \dd \phi^j_{i,I} .
\]
Each factor in the integrand is given by
\begin{gather*}
 z_i^\text{vec} = [-\epsilon_{12}]^{k_i} \prod_{j=0}^{p-1} \Bigg [ \prod_{I=1}^{k_i^j} \frac{1}{P_i^j\big(\phi_{i,I}^j\big) \widetilde{P}_i^j\big(\phi_{i,I}^j+\epsilon_{12}\big)}
 \nonumber \\
 \hphantom{z_i^\text{vec} = [-\epsilon_{12}]^{k_i} \prod_{j=0}^{p-1} \Bigg [ \prod_{I=1}^{k_i^j}}{}
 \times
 \frac{\prod_{I \neq J}^{k_i^j} \big[\phi^j_{i,I} - \phi_{i,J}^j\big] \big[\phi^j_{i,I} - \phi_{i,J}^j - \epsilon_{12}\big]}{\prod_{J = 1,\ldots,k_i^{j+1}}^{I = 1,\ldots, k_i^j} \big[\phi_{i,I}^j - \phi_{i,J}^{j+1} - \epsilon_1\big] \prod_{J = 1,\ldots,k_i^{j-1}}^{I = 1,\ldots, k_i^j} \big[\phi_{i,I}^j - \phi_{i,J}^{j-1} - \epsilon_2\big]} \Bigg]
 , \\
 z_{e\colon i \to i'}^\text{bf} = \prod_{j=0}^{p-1} \Bigg[ \prod_{I = 1}^{k_{i'}} P_i^j\big(\phi_{i',I}^j + m_e\big) \prod_{I = 1}^{k_i} \widetilde{P}_{i'}^j\big(\phi_{i,I}^j - m_e + \epsilon_{12}\big)
 \nonumber \\
\hphantom{z_{e\colon i \to i'}^\text{bf} = \prod_{j=0}^{p-1}}{}
 \times \frac{\prod_{J = 1,\ldots,k_i^{j+1}}^{I = 1,\ldots, k_{i'}^j} \!\!\big[\phi_{i',I}^j - \phi_{i,J}^{j+1} + m_e - \epsilon_{1}\big]\! \prod_{J = 1,\ldots,k_i^{j-1}}^{I = 1,\ldots, k_{i'}^j} \!\!\big[\phi_{i',I}^j - \phi_{i,J}^{j-1} + m_e - \epsilon_{2}\big]}{\prod^{I = 1, \ldots, k_{i'}^j}_{J = 1,\ldots, k_i^j}\!\! \big[\phi^j_{i',I} - \phi_{i,J}^j + m_e\big] \big[\phi^j_{i',I} - \phi_{i,J}^j + m_e - \epsilon_{12}\big]}
 \Bigg]
 , \\
 z_i^\text{f} = \prod_{j=0}^{p-1} \prod_{I = 1}^{k_i^j} P_{i}^{\text{f},j}\big(\phi_{i,I}^j\big)
 , \qquad
 z_i^\text{af} = \prod_{j=0}^{p-1} \prod_{I = 1}^{k_i^j} \widetilde{P}_{i}^{\text{f},j}\big(\phi_{i,I}^j + \epsilon_{12}\big) .
\end{gather*}
The gauge and matter polynomials are defined as
\begin{alignat*}{3}
& P_i^j(x) = P_{\mathbf{N}_i^j}(x) = \prod_{\alpha = 1}^{n_i^j} \big[x - a_{i,\alpha}^j\big] , \qquad&&
 \widetilde{P}_i^j(x) = \widetilde{P}_{\mathbf{N}_i^j}(x) = \prod_{\alpha = 1}^{n_i^j} \bigl[- x + a_{i,\alpha}^j\bigr] , & \\
 & P_i^{\text{f},j}(x) = P_{\mathbf{M}_i^j}(x) = \prod_{f = 1}^{n_i^{\text{f},j}} \big[x - m_{i,f}^j\big] , \qquad&&
 \widetilde{P}_i^{\text{f},j}(x) = \widetilde{P}_{\widetilde{M}_i^j}(x) = \prod_{f = 1}^{n_i^{\text{af},j}} \bigl[- x + \widetilde{m}_{i,f}^j\bigr] .&
\end{alignat*}

In addition, we may incorporate a defect operator using the observable sheaves as before.
With a one-dimensional bundle $\mathbf{x}$ with the character $\operatorname{ch} \mathbf{x} = {\rm e}^x$, we define the $\mathscr{Y}$-functions
\begin{gather*}
 \mathscr{Y}_i^j(x) = \mathbb{I}\big[\mathbf{Y}_i^{j\vee} \mathbf{x}\big]
 = \prod_{\alpha = 1}^{n_i^j} \big[x - a_{i,\alpha}^j\big] \frac{\prod_{I = 1,\ldots,k_i^{j+1}}\big[x - \phi_{i,I}^{j+1} - \epsilon_1\big] \prod_{I = 1,\ldots,k_i^{j-1}} \big[x - \phi_{i,I}^{j-1} - \epsilon_2\big]}{\prod_{I = 1,\ldots,k_i^j}\big[x - \phi_{i,I}^j\big]\big[x - \phi_{i,I}^j - \epsilon_{12}\big]} .
\end{gather*}
We can similarly consider the dual $\mathscr{Y}$-function, $\widetilde{\mathscr{Y}}_i^j(x) = \mathbb{I}\big[\mathbf{x}^\vee \mathbf{Y}_i^{j}\big]$.

The total partition function is given by summing over all the topological sectors as before~\eqref{eq:tot_pf_quiv}.
In this case, the expression is almost parallel with that for quiver gauge theory, and we have the instanton counting parameter for each node, $j = 0,\ldots,p-1$,
\[
 Z = \sum_{\underline{k}} \underline{\mathfrak{q}}^{\underline{k}} Z_\gamma .
\]
These counting parameters can be converted to those associated with the second and first Chern classes of the tautological bundle over the ALE space~\cite{Fucito:2004ry}.

\subsubsection[A\_1 quiver: SQCD]{$\boldsymbol{A_1}$ quiver: SQCD}

The first example of the orbifold instanton partition function is SQCD on $\mathbb{C}^2/\mathbb{Z}_p$ ($A_1$ quiver on $\mathbb{C}^2/\mathbb{Z}_p$).
Applying the generic formula to this case, we obtain the instanton partition function
\begin{gather*}
 Z_\gamma = \frac{[-\epsilon_{12}]^{k}}{\underline{k}!} \oint_{\mathbb{T}_\mathbf{K}} \dd{\underline{\phi}} \prod_{j=0}^{p-1} \Bigg[ \prod_{I=1}^{k_i} \frac{P^{\text{f},j}\big(\phi_I^j\big) \widetilde{P}^{\text{f},j}\big(\phi_I^j+\epsilon_{12}\big)}{P^j(\phi_I^j) \widetilde{P}^j\big(\phi_I^j+\epsilon_{12}\big)}
 \nonumber \\
 \hphantom{Z_\gamma = \frac{[-\epsilon_{12}]^{k}}{\underline{k}!} \oint_{\mathbb{T}_\mathbf{K}}}{}
 \times
 \frac{\prod_{I \neq J}^{k_i} \big[\phi^j_I - \phi_J^j\big] \big[\phi^j_I - \phi_J^j - \epsilon_{12}\big]}{\prod_{J = 1,\ldots,k^{j+1}}^{I = 1,\ldots, k^j} \big[\phi_I^j - \phi_J^{j+1} - \epsilon_1\big] \prod_{J = 1,\ldots,k^{j-1}}^{I = 1,\ldots, k^j} \big[\phi_I^j - \phi_J^{j-1} - \epsilon_2\big]} \Bigg].
\end{gather*}
Evaluating the residues of this contour integral, we obtain the combinatorial form of the instanton partition function for $\mathbb{C}^2/\mathbb{Z}_p$~\cite{Belavin:2011pp,Bonelli:2011jx,Bonelli:2011kv, Fucito:2004ry,Fujii:2005dk}.

\subsubsection[widehat A\_0 quiver]{$\boldsymbol{\widehat{A}_0}$ quiver}

We consider $\widehat{A}_0$ quiver gauge theory on the orbifold $\mathbb{C}^2/\mathbb{Z}_p$ similarly to Section~\ref{sec:Ar^_quiver}.
In this case, the contour integral form of the instanton partition function is given as follows:
\begin{gather*}
 Z_\gamma = \frac{1}{\underline{k}!} \frac{[-\epsilon_{12}]^{k}}{[-\epsilon_3]^k[\epsilon_4]^k} \oint_{\mathbb{T}_\mathbf{K}} \dd{\underline{\phi}} \prod_{j=0}^{p-1} \Bigg[ \prod_{I=1}^{k_i} \frac{P^j\big(\phi_I^j-\epsilon_3\big) \widetilde{P}^j\big(\phi_I^j+\epsilon_{4}\big)}{P^j\big(\phi_I^j\big) \widetilde{P}^j\big(\phi_I^j+\epsilon_{12}\big)}
 \prod_{I \neq J}^{k_i} \frac{\big[\phi^j_{IJ}\big] \big[\phi^j_{IJ} - \epsilon_{12}\big]}{\big[\phi^j_{IJ}-\epsilon_3\big] \big[\phi^j_{IJ} + \epsilon_{4}\big]}
 \nonumber \\
 \hphantom{Z_\gamma = \frac{1}{\underline{k}!} \frac{[-\epsilon_{12}]^{k}}{[-\epsilon_3]^k[\epsilon_4]^k} \oint_{\mathbb{T}_\mathbf{K}}}{}
\times
 \prod_{\substack{I = 1,\ldots, k^j \\ J = 1,\ldots,k^{j+1}}} \frac{\big[\phi_I^j - \phi_J^{j+1} - \epsilon_{13}\big]}{\big[\phi_I^j - \phi_J^{j+1} - \epsilon_1\big]}
 \prod_{\substack{I = 1,\ldots, k^j \\ J = 1,\ldots,k^{j-1}}} \frac{\big[\phi_I^j - \phi_J^{j-1} - \epsilon_{23}\big]}{\big[\phi_I^j - \phi_J^{j-1} - \epsilon_2\big]}
 \Bigg]
 \nonumber \\
\hphantom{Z_\gamma}{}
 =
 \frac{q_3^{nk +\sum_{j=0}^{p-1} k_j k_{j+1} }}{\underline{k}!} \frac{[-\epsilon_{12}]^{k}}{[-\epsilon_3]^k[\epsilon_4]^k} \\
 \hphantom{Z_\gamma=}{}\times
 \oint \dd{\underline{\phi}} \prod_{j=0}^{p-1} \Bigg[ \prod_{\substack{I=1,\ldots,k_i \\ \alpha = 1,\ldots,n_i}}
 \mathscr{S}_{34}\big(\phi_I^j - a^j_\alpha\big)
 \frac{\prod_{J = 1,\ldots,k^{j+1}}^{I = 1,\ldots, k^j} \mathscr{S}_{34}\big(\phi_I^j-\phi_J^{j+1}-\epsilon_1\big)}{\prod_{I < J}^{k_i} \mathscr{S}_{34}\big(\phi_{IJ}^j\big) \mathscr{S}_{34}\big(\phi_{IJ}^j-\epsilon_{12}\big)}
 \Bigg] ,
\end{gather*}
where we use the notation~\eqref{eq:epsilon34} for the adjoint mass parameter.
Compared to $\widehat{A}_{r-1}$ quiver theory on $\mathbb{C}^2$ given in~\eqref{eq:Ar-1^_quiver}, we have the same measure factor up to the replacement of $(\epsilon_{1,2})$ with $(\epsilon_{3,4})$ although the source term is different.\footnote{More precisely, we have the prefactor quadratically depending on $\underline{k} = (k_i)_{i=0,\ldots,p-1}$ for the 5d/6d convention, which cannot be absorbed by redefinition of the counting parameter.}
This similarity is naturally explained through their geometric realization in eight dimensions, $\mathbb{C}^2 \times\big(\mathbb{C}^2/\mathbb{Z}_r\big)$ and $\big(\mathbb{C}^2/\mathbb{Z}_p\big) \times \mathbb{C}^2$.
See Section~\ref{sec:geometry}.

\section{Double quiver theory}\label{sec:doubld_quiver_theory}

In Section~\ref{sec:orb_inst}, we have seen that the orbifold instanton partition function is concisely described using the $q$-Cartan matrix of the McKay--Nakajima quiver $\Upsilon$.
Let us rewrite the generic formula for the vector and hypermultiplet contributions (see \eqref{eq:vect_contribution_upsilon_full} and \eqref{eq:hyp_contribution_upsilon_full})
\begin{alignat}{3}
 &\mathbf{V}_i = \mathbf{Y}^\vee_i \tilde{\mathbf{c}}_\Upsilon \mathbf{Y}_i , \qquad&&
 \mathbf{H}_{e\colon i \to i'} = - \mathbf{M}_e \mathbf{Y}^\vee_i \tilde{\mathbf{c}}_\Upsilon \mathbf{Y}_j , &\nonumber\\
& \mathbf{H}_i = - \mathbf{M}_i^\vee \tilde{\mathbf{c}}_\Upsilon \mathbf{Y}_i , \qquad&&
 \widetilde{\mathbf{H}}_i = - \mathbf{Y}_i^\vee \tilde{\mathbf{c}}_\Upsilon \widetilde{\mathbf{M}}_i ,&
 \label{eq:double_quiver_V_H}
\end{alignat}
for $i, i' \in \Gamma_0$, where $\mathbf{c}_\Upsilon$ is the $q$-Cartan matrix with respect to $\Upsilon$-structure and its inverse $\tilde{\mathbf{c}}_\Upsilon$, and the observable sheaf is given by
\[
 \mathbf{Y}_i = \mathbf{N}_i - \mathbf{c}_\Upsilon \mathbf{K}_i .
\]
We then define the $\mathscr{Y}$-function to impose the codimension-four defect
\[
 \mathscr{Y}_i^j(x) = \mathbb{I}\big[\mathbf{Y}_i^{j\vee} \mathbf{x}\big] , \qquad
 \widetilde{\mathscr{Y}}_i^j(x) = \mathbb{I}\big[\mathbf{x}^\vee \mathbf{Y}_i^j\big] .
\]
From this point of view, it is natural to consider generic $q$-Cartan matrix in this construction even though there is no direct geometric realization.

We call the complex surface characterized by $\Upsilon$-quiver the $\Upsilon$ surface $\mathcal{S}_\Upsilon$, which becomes the ALE space when $\Upsilon$ is an $\widehat{{\rm ADE}}$ quiver.
We may formally consider the case with a finite type $\Upsilon$ quiver as well.
Then, we call $\Gamma$-quiver gauge theory on the $\Upsilon$-surface the \emph{double quiver gauge theory} of type $(\Upsilon,\Gamma)$.
Hence, the corresponding instanton partition function is given by the equivariant integral over the quiver variety of type $\Upsilon$.

The formalism of double quiver theory is still applicable to fractional quiver, which may describe non-simply-laced algebras, by replacing the $q$-Cartan matrix with its symmetrization~\cite{Kimura:2017hez}.
In this case, the partition function is given by the integral over the fractionalization of quiver variety~\cite{KPfractional}.
See also \cite{Nakajima:2019olw}.

The double quiver theory has another realization in particular for $(\Upsilon,\Gamma) = \big(\widehat{A}_1,\Gamma\big)$ as $\Gamma$-quiver supergroup gauge theory~\cite{Kimura:2020jxl, Kimura:2019msw}.
This is due to the fact that supergroup gauge theory is obtained through the analytic continuation of $\widehat{A}_1$ quiver theory~\cite{Dijkgraaf:2016lym}.
Hence, the supergroup gauge theory is in general obtained from quiver gauge theory on $\mathcal{S}_{\widehat{A}_1} = \mathbb{C}^2/\mathbb{Z}_2$.
We remark that a similar connection between supergroup structure and the $\mathbb{Z}_2$ quotient space is discussed for lower dimensional theory:
A supergroup analog of Chern--Simons theory, a.k.a., ABJ(M) theory, defined on $\mathbb{S}^3$ is obtained from the analytic continuation of the Chern--Simons theory on the lens space $\mathbb{S}^3/\mathbb{Z}_2$~\cite{Drukker:2009hy,Marino:2009jd}.
Moreover, it has been known that the instanton partition function on $\mathcal{S}_{\widehat{A}_1} = \mathbb{C}^2/\mathbb{Z}_2$ is identified with the super Liouville conformal block~\cite{Belavin:2011pp,Bonelli:2011jx,Bonelli:2011kv,Nishioka:2011jk}, which also provides a link between the supersymmetric structure and the $\mathbb{Z}_2$ quotient space.

\subsection{Instanton partition function}

In order to describe double quiver theory $(\Upsilon,\Gamma)$ where $\Upsilon = (\Upsilon_0,\Upsilon_1)$ and $\Gamma = (\Gamma_0,\Gamma_1)$, we consider the associated $q$-Cartan matrices,
\begin{gather*}
 \mathbf{c}_{\Upsilon,ii'} = (1 + \mathbf{Q}_{12}) \delta_{ii'} - \sum_{e \in \Upsilon_1\colon i \to i'} \mathbf{M}_e^{\vee} - \sum_{{\rm e}^\vee \in \Upsilon_1^\vee: i' \to i} \mathbf{M}_e \mathbf{Q}_{12} , \qquad i, i' \in \Upsilon_0 , \\
 \mathbf{c}_{\Gamma,jj'} = (1 + \mathbf{Q}_{34}) \delta_{jj'} - \sum_{e \in \Gamma_1\colon j \to j'} \mathbf{M}_e^{\vee} - \sum_{{\rm e}^\vee \in \Gamma_1^\vee: j' \to j} \mathbf{M}_e \mathbf{Q}_{34} , \qquad i, i' \in \Gamma_0 .
\end{gather*}
As mentioned in the beginning of Section~\ref{sec:quiver_gauge_theory}, all the edges have their duals with opposite orientation, $e\colon i \to i'$ and ${\rm e}^\vee\colon i' \to i$.
We remark that the relation $q_{12} = q_{34}^{-1}$ comes from the Calabi--Yau condition with respect to $\mathbb{C}^4$ (see~\eqref{eq:CY4_cond}), $q_1 q_2 q_3 q_4 = 1$, and hence, in the character notation, we have $c_{\Upsilon}^\vee = q_{12}^{-1} c_\Upsilon^\text{T}$ and $c_{\Gamma}^\vee = q_{34}^{-1} c_\Gamma^\text{T} = q_{12} c_\Gamma^\text{T}$.

From the vector and hypermultiplet contributions shown in \eqref{eq:double_quiver_V_H}, we may write the perturbative part and the instanton part using the $q$-Cartan matrices defined above
\begin{subequations}
\begin{align}
&
 \bigg[ \sum_{i \in \Gamma_0} \big( \mathbf{V}_i + \mathbf{H}_i + \widetilde{\mathbf{H}}_i \big) + \sum_{e \in \Gamma_1} \mathbf{H}_{e} \bigg]_{\mathbf{K} \to 0}
 = \sum_{\substack{i,i' \in \Gamma_0 \\ j,j' \in \Upsilon_0}} \mathbf{N}_i^{j\vee} \tilde{\mathbf{c}}_{\Upsilon,jj'} \mathbf{c}^{+\vee}_{\Gamma,ii'} \mathbf{N}_{i'}^{j'}
 , \label{eq:double_quiver_pert} \\
 &
 \bigg[ \sum_{i \in \Gamma_0} \big( \mathbf{V}_i + \mathbf{H}_i + \widetilde{\mathbf{H}}_i \big) + \sum_{e \in \Gamma_1} \mathbf{H}_{e} \bigg]_\text{inst}
 = \sum_{\substack{i,i' \in \Gamma_0 \\ j,j' \in \Upsilon_0}} \Bigl[ - \big( \mathbf{N}_i^{j\vee} \delta_{j,j'} \mathbf{c}^{+\vee}_{\Gamma,ii'} \mathbf{K}_{i'}^{j'} + \mathbf{N}_i^{j} \delta_{j,j'} \mathbf{c}^-_{\Gamma,ii'} \mathbf{K}_{i'}^{j'\vee} \big)
 \nonumber \\
 & \qquad{}
 + \mathbf{K}_i^{j\vee} \mathbf{c}_{\Upsilon,jj'}^\dag \mathbf{c}^{+\vee}_{\Gamma,ii'} \mathbf{K}_{i'}^{j'} + \mathbf{M}_i^{j\vee} \delta_{j,j'} \delta_{i,i'} \mathbf{K}_{i'}^{j'} + \mathbf{Q}_{12}^\vee \mathbf{K}_i^{j\vee} \delta_{j,j'} \delta_{i,i'} \widetilde{\mathbf{M}}_{i'}^{j'} \Bigr] ,
 \label{eq:double_quiver_total}
\end{align}
\end{subequations}
where $\mathbf{c}_\Gamma^\pm$ are the half $q$-Cartan matrices for quiver $\Gamma$~\eqref{eq:half_Cartan_mat}.
Denoting the topological data by $\gamma = (n,k)$, we obtain the instanton partition function by applying the index functor
\[
 Z_\gamma = \mathbb{I}\bigg[ \sum_{i \in \Gamma_0} \big( \mathbf{V}_i + \mathbf{H}_i + \widetilde{\mathbf{H}}_i \big) + \sum_{e \in \Gamma_1} \mathbf{H}_{e} \bigg]_\text{inst} .
\]
The total partition function is then given by summation over the topological sectors
\[
 Z = \sum_{\underline{k}} \underline{\mathfrak{q}}^{\underline{k}} Z_\gamma , \qquad
 \underline{\mathfrak{q}}^{\underline{k}} = \prod_{\substack{i \in \Gamma_0 \\ j \in \Upsilon_0}} (\mathfrak{q}_i^j)^{k_i^j} .
\]
Even though it is not difficult to write down the contour integral formula explicitly, it does not seem useful enough.
We instead discuss several examples in the following.

\subsection{Examples}

\subsubsection[widehat A\_\{p-1\} surface]{$\boldsymbol{\widehat{A}_{p-1}}$ surface}

We first consider the $\widehat{A}_{p-1}$ surface, which gives rise to the orbifold surface of type $A$, $\mathcal{S}_{\widehat{A}_{p-1}} = \mathbb{C}^2/\mathbb{Z}_p$.
In this case, the $q$-Cartan matrix is given by \eqref{eq:qCartan_Ap-1}, and we obtain the instanton partition function on $\mathbb{C}^2/\mathbb{Z}_p$ as discussed in Section~\ref{sec:orb_inst}.

\subsubsection[widehat D\_p surface]{$\boldsymbol{\widehat{D}_p}$ surface}

We consider the $\widehat{D}_p$ surface, which corresponds to the orbifold of type $D$, $\mathcal{S}_{\widehat{D}_p} = \mathbb{C}^2/\mathbb{BD}_{p-2}$, where we denote the binary dihedral group of degree $r$ by $\mathbb{BD}_{r}$.
The $q$-Cartan matrix is given as follows in the character notation:
\begin{gather*}
 {c}_\Upsilon =
 \begin{pmatrix}
 1 + q_{12} & & - \mu_0 & & & & \\
 & 1 + q_{12} & - \mu_1 & & & & & \\
 - \mu_0^{-1} q_{12} & - \mu_1^{-1} q_{12} & 1 + q_{12} & - \mu_2 & & & & \\
 && - \mu_2^{-1} q_{12} & 1 + q_{12} & - \mu_3 &&& \\
 &&& - \mu_3^{-1} q_{12} & \ddots & \ddots && \\
 &&&& \ddots & 1 + q_{12} & - \mu_{p-2} & - \mu_{p-1} \\
 &&&&& - \mu_{p-2}^{-1} q_{12} & 1 + q_{12} & \\
 &&&&& - \mu_{p-1}^{-1} q_{12} && 1 + q_{12}
 \end{pmatrix} .
\end{gather*}
We denote the bifundamental mass parameters by $(\mu_i)_{i = 0,\ldots,p-1}$.
The determinant of the $q$-Cartan matrix of type $\widehat{D}_p$ is given by
\[
 \det c_\Upsilon = (1 + q_{12}) \big(1 - q_{12}^2\big) \big(1 - q_{12}^{p-2}\big) ,
\]
which does not depend on the bifundamental mass parameters.

Hereafter, we simply set all the bifundamental masses to be trivial $\mu_i = 1$ by gauge fixing,\footnote{These bifundamental mass parameters in the $q$-Cartan matrix $c_\Upsilon$ are originally interpreted as the spacetime parameters. This gauge fixing is equivalent to redefinition of the irreducible representation character of~$\Upsilon$.}
and we have only a single deformation parameter~$q_{12}$.
This means that we cannot impose generic $\Omega$-background parameter for the D-type quotient:
The binary dihedral group is in general a non-Abelian group, which involves a generator exchanging two coordinates of~$\mathbb{C}^2$.
Hence, it is not compatible with the full Cartan torus of~${\rm GL}(\mathbf{Q})$ denoted by~$\mathbb{T}_\mathbf{Q}$.

Let us consider the case $p = 4$ with the quiver structure
$\Upsilon = \dynkin[label]{D}[1]{4}$.
The $q$-Cartan matrix is in this case given by
\[
 c_\Upsilon =
 \begin{pmatrix}
 1 + q_{12} & & -1 & & \\
 & 1 + q_{12} & -1 && \\
 - q_{12} & - q_{12} & 1 + q_{12} & -1 & -1 \\
 && - q_{12} & 1 + q_{12} & \\
 && - q_{12} && 1 + q_{12}
 \end{pmatrix} .
\]
Then, we obtain the contour integral formula as follows:
\begin{gather*}
 Z_\gamma = \frac{[-\epsilon_{12}]^k}{\underline{k}!} \oint_{\mathbb{T}_\mathbf{K}} \dd{\underline{\phi}} \prod_{\substack{j = 0,\ldots,4 \\ I = 1,\ldots,k^j}} \frac{1}{P^j\big(\phi_I^j\big) \widetilde{P}\big(\phi_I^j + \epsilon_{12}\big)}
 \nonumber \\
 \times
 \frac{\prod_{j=0}^4 \prod_{I \neq J}^{k^j} \big[\phi^j_{IJ}\big]\big[\phi^j_{IJ}-\epsilon_{12}\big]}{\prod_{j = 0, 1} \prod_{J = 1,\ldots,k^j}^{I = 1,\ldots,k^2} \big[\phi_I^2 - \phi_J^j\big] \big[\phi_J^j - \phi_I^2 - \epsilon_{12}\big] \prod_{j = 3,4} \prod_{J = 1,\ldots,k^j}^{I = 1,\ldots,k^2} \big[\phi_J^j - \phi_I^2\big] \big[\phi_I^2 - \phi_J^j - \epsilon_{12}\big]} .
\end{gather*}
We remark that a similar formula has been obtained by equivariant integration over quiver variety in general~\cite{Moore:1997dj}.
In this case, the pole structure of the integrand is largely different from the type $A$ case, and the combinatorial form of the instanton partition function is not available at this moment.

\subsubsection[A\_1 surface]{$\boldsymbol{A_1}$ surface}\label{sec:A1_surface}

We then consider the case $\Upsilon = A_1$ with the $q$-Cartan matrix,
\[
 \mathbf{c}_\Upsilon = 1 + \mathbf{Q}_{12} .
\]
In this case, the partition function for the topological data $\gamma = (n,k)$ is given as follows:
\begin{gather}
 Z_\gamma = \frac{[-\epsilon_{12}]^k}{k!} \oint_{\mathbb{T}_\mathbf{K}} \prod_{I = 1}^k \frac{\dd{\phi}_I}{2 \pi \iota} \frac{1}{P(\phi_I) \widetilde{P}(\phi_I + \epsilon_{12})} \prod_{I \neq J}^k [\phi_{IJ}][\phi_{IJ} - \epsilon_{12}],
 \label{eq:A1A1_inst_part}
\end{gather}
where we have the gauge polynomials $P(x)$ and $\widetilde{P}(x)$ as before.
The difference from the LMNS formula of type $\widehat{A}_0$ shown in~\eqref{eq:LMNS_formula} is the denominator in the measure term, which corresponds to the contribution of loop edges in $\widehat{A}_0$ quiver.
Hence, the residue structure becomes much simpler in this case:
Each $(\phi_I)_{I = 1,\ldots,k}$ chooses one of $(a_\alpha)_{\alpha = 1,\ldots,n}$.
These $n \choose k$ choices correspond to the fixed points of the cotangent bundle of Grassmannian ${\rm Gr}(k,n)$, which is the quiver variety of type $A_1$, $\mathfrak{M}_{\gamma} = \{ I \in \operatorname{Hom}(\mathbf{N},\mathbf{K}), J \in \operatorname{Hom}(\mathbf{K},\mathbf{N}) \mid IJ = 0 \} / \!\! / \mathrm{GL}(\mathbf{K})$.\footnote{The constraint appearing in the definition of $\mathfrak{M}_\gamma$ in this case is obtained from the potential $\mathcal{W} = \tr \phi IJ$, whose derivative gives rise to $\partial \mathcal{W} / \partial \phi = IJ$.}
The Grassmannian becomes empty for $k > n$ where the partition function becomes trivial.
For the case $\Upsilon = A_p$, the quiver variety becomes the cotangent bundle of the flag manifold.

Applying the general formula of the $\mathscr{Y}$-function to the current case, we have
\[
 \mathscr{Y}(x) = \prod_{\alpha=1}^n [x - a_\alpha] \prod_{I = 1}^k [x-\phi_I]^{-1} [x - \phi_I - \epsilon_{12}]^{-1}.
\]
Hence, denoting the topological data $\gamma = (n,n',k)$ for the vector space $\mathbf{N}' = \mathbb{C}^{n'}$ with $\operatorname{ch} \mathbf{N}' = \sum_{\beta = 1}^{n'} {\rm e}^{b_\beta}$, the instanton partition function is given by
\begin{align*}
 Z_\gamma & = \frac{[-\epsilon_{12}]^k}{k!} \oint_{\mathbb{T}_\mathbf{K}} \dd{\underline{\phi}} \mathscr{Y}_{\mathbf{N}'} \prod_{I = 1}^k \frac{1}{P(\phi_I) \widetilde{P}(\phi_I + \epsilon_{12})} \prod_{I \neq J}^k [\phi_{IJ}][\phi_{IJ} - \epsilon_{12}] \nonumber \\
 & = \frac{[-\epsilon_{12}]^k}{k!} \oint_{\mathbb{T}_\mathbf{K}} \dd{\underline{\phi}}
 \frac{\prod^{\alpha = 1,\ldots,n}_{\beta = 1,\ldots,n'} [b_\beta - a_\alpha] \prod_{I \neq J}^k [\phi_{IJ}][\phi_{IJ}-\epsilon_{12}]}{\prod_{I = 1,\ldots,k}^{\alpha = 1,\ldots,n} [\phi_I - a_\alpha] [a_\alpha - \phi_I - \epsilon_{12}] \prod_{I = 1,\ldots,k}^{\beta = 1,\ldots,n'} [b_\beta - \phi_I][b_\beta - \phi_I - \epsilon_{12}]} .
\end{align*}
Now the symmetry between $(a_\alpha)_{\alpha = 1,\ldots,n}$ and $(b_\beta)_{\beta = 1,\ldots,n'}$ is obvious.

We then add the adjoint matter contribution.
Namely, we consider the case $(\Upsilon,\Gamma) = \big(A_1,\widehat{A}_0\big)$.
The partition function is given by
\begin{align*}
 Z_\gamma & = \frac{1}{k!} \frac{[-\epsilon_{12}]^k}{[-\epsilon_3]^k[\epsilon_4]^k} \oint_{\mathbb{T}_\mathbf{K}} \prod_{I = 1}^k \frac{\dd{\phi}_I}{2 \pi \iota} \frac{P(\phi_I - \epsilon_3) \widetilde{P}(\phi_I + \epsilon_{4})}{P(\phi_I) \widetilde{P}(\phi_I + \epsilon_{12})} \prod_{I \neq J}^k \frac{[\phi_{IJ}][\phi_{IJ} - \epsilon_{12}]}{[\phi_{IJ} - \epsilon_3][\phi_{IJ} + \epsilon_{4}]}
 \nonumber \\
 & = \frac{q_3^{nk}}{k!} \frac{[-\epsilon_{12}]^k}{[-\epsilon_3]^k[\epsilon_4]^k} \oint_{\mathbb{T}_\mathbf{K}} \prod_{I = 1}^k \frac{\dd{\phi}_I}{2 \pi \iota} \prod_{\alpha = 1}^n \mathscr{S}_{34}(\phi_I - a_\alpha) \prod_{I < J}^k \mathscr{S}_{34}(\phi_{IJ})^{-1} \mathscr{S}_{34}(\phi_{IJ} + \epsilon_{34})^{-1},
\end{align*}
where we apply the notation \eqref{eq:epsilon34} for the adjoint mass parameter.
In this case, even though the source term is different, the measure factor is equivalent to the original LMNS formula \eqref{eq:LMNS_formula} for $(\Upsilon,\Gamma) = \big(\widehat{A}_0,A_1\big)$ by replacing $\epsilon_{1,2} \leftrightarrow \epsilon_{3,4}$.

\subsubsection[BC\_2 surface]{$\boldsymbol{BC_2}$ surface}

We consider the $BC_2$ quiver, $\Upsilon = \dynkin[label]{B}{2}$, which is the minimal example beyond the ${\rm ADE}$ quivers.
The $q$-Cartan matrix is in this case given by~\cite{Kimura:2017hez}
\[
 c_\Upsilon =
 \begin{pmatrix}
 1 + q_1^2 q_2 & - \mu \\
 -\mu^{-1} q_{12} (1 + q_1) & 1 + q_{12}
 \end{pmatrix}
\]
with the determinant
\[
 \det c_\Upsilon = 1 + q_1^3 q_2
 \xrightarrow{q_{1,2} \to 1} 2 .
\]
Moreover, we define the $q$-deformation of the symmetrized Cartan matrix
\[
 b_\Upsilon =
 \begin{pmatrix}
 (1 + q_1)\big(1 + q_1^2 q_2\big) & - \mu (1 + q_1) \\
 -\mu^{-1} q_{12} (1 + q_1) & 1 + q_{12}
 \end{pmatrix} .
\]
The double quiver theory associated with the surface $\mathcal{S}_\Upsilon$ is defined by replacing the $q$-Cartan matrix with its symmetrization.
For example, the contour integral form of the instanton partition function for $(\Upsilon,\Gamma) = (BC_2,A_1)$ is given by
\begin{gather*}
 Z_\gamma = \frac{c_{\underline{k}}}{k_1! k_2!} \oint_{\mathbb{T}_\mathbf{K}} \dd{\underline{\phi}}
 \prod_{i=1}^2 \prod_{I = 1}^{k_i} \frac{1}{P_i\big(\phi_{I}^i\big) \widetilde{P}_i\big(\phi_{I}^i + \epsilon_{12}\big) }
 \nonumber \\
 \times
 \frac{\prod_{I \neq J}^{k_1} \big[\phi^1_{IJ}\big] \big[\phi^1_{IJ} - \epsilon_{1,1^2 2, 1^3 2\big]} \prod_{I \neq J}^{k_2} \big[\phi^2_{IJ}\big] \big[\phi^2_{IJ} - \epsilon_{1 2}\big] }{ \prod^{I = 1,\ldots,k_1}_{J = 1,\ldots,k_2} \big[\phi^2_J - \phi^1_I - m\big]\big[\phi^2_J - \phi^1_I - m - \epsilon_1\big]\big[\phi^1_I - \phi^2_J + m - \epsilon_{12}\big]\big[\phi^1_I - \phi^2_J + m - \epsilon_{1^2 2}\big]},
\end{gather*}
where $P_i(\phi)$ is the gauge polynomial for each node $i = 1,2$, and we denote the constant term by
\[
 c_{\underline{k}} = [-\epsilon_{1,1^2 2, 1^3 2}]^{k_1} [-\epsilon_{12}]^{k_2} = ([-\epsilon_{1}][-\epsilon_{1^2 2}][-\epsilon_{1^3 2}])^{k_1} ([-\epsilon_{12}])^{k_2} .
\]
This instanton partition function computes the equivariant volume of the fractional quiver variety of type $BC_2$~\cite{KPfractional}.

\subsection{Geometric realization}\label{sec:geometry}

The double quiver theory has a natural geometric realization in eight dimensions.
We start with the Hanany--Witten configuration of eight supercharge quiver gauge theory of type $A$ and $\widehat{A}$ on~$\mathbb{C}^2$ as follows:
\begin{align}
 \begin{tabular*}{.75\textwidth}{@{\extracolsep{\fill}}ccccccccccc} \toprule
 (IIA) &0&1&2&3&4&5&6&7&8&9 \\ \midrule
 NS5 &--&--&--&--&--&--&&& \\
 D4 &--&--&--&--&&&--&&& \\
 D6 &--&--&--&--&&&&--&--&-- \\\toprule
 \end{tabular*}
 \label{eq:IIAsetup}
\end{align}
where $\mathbb{C}^2$ corresponds to 0123-directions.
NS5 branes are aligned in 6-direction that suspends~D4 branes within finite intervals.
D6 branes are used to consider the hypermultiplet in the fundamental representation.
The 6-direction is cyclic for affine~$\widehat{A}$ quiver, while it becomes infinite for finite~$A$ quiver.
Hence, we may consider the T-duality to the 6-direction for the case~$\widehat{A}_{r-1}$, which geometrizes~$r$ NS5 branes to the $r$-centered Taub-NUT space,
\begin{align}
 \begin{tabular*}{.75\textwidth}{@{\extracolsep{\fill}}ccccccccccc} \toprule
 (IIB) &0&1&2&3&4&5&6&7&8&9 \\ \midrule
 TN &&&&&&&--&--&--&-- \\
 D3 &--&--&--&--&&&&&& \\
 D7 &--&--&--&--&&&--&--&--&-- \\\toprule
 \end{tabular*}
 \label{eq:IIBsetup}
\end{align}
Taking the radius of the $\mathbb{S}^1$ fibration in the Taub-NUT space to infinity, we obtain the ALE space of type $A$.
This process is similarly applied to the case $\Gamma = \widehat{{\rm ADE}}$ to obtain the ADE singularity in 6789-directions.\footnote{The brane realization is discussed for D-type~\cite{Hanany:1999sj, Kapustin:1998fa} and E-type~\cite{Kimura:2019gon}.}
Replacing 0123-directions with the ALE space, we obtain the double quiver theory $(\Upsilon,\Gamma)$ where both $\Upsilon$ and $\Gamma$ are of affine type.
Therefore, we have a geometric realization of $(\Upsilon,\Gamma)$-theory on $(\mathbb{C}_1 \times \mathbb{C}_2)/\Upsilon \times (\mathbb{C}_3 \times \mathbb{C}_4)/\Gamma$, which explains that exchanging $\Upsilon$ and $\Gamma$ is equivalent to that for $\epsilon_{1,2}$ and $\epsilon_{3,4}$.

Moreover, it is also possible to add D($-1$) branes and D3 branes extended in 6789-directions (denoted by D3') in the type IIB setup~\eqref{eq:IIBsetup}.
In this context, D($-1$) branes are interpreted as the instantons, and the D3$'$ brane plays a role of the defect brane, which gives rise to the codimension-four defect in $\Gamma$-quiver gauge theory on $\mathbb{C}^2/\Upsilon$ described by the $\mathscr{Y}$-functions.
Denoting the surface associated with $\Upsilon$ and $\Gamma$ quivers by $\mathcal{S}_\Gamma$ and $\mathcal{S}_\Upsilon$, we have the configuration
\begin{align}
 \begin{tabular*}{.75\textwidth}{@{\extracolsep{\fill}}ccccccccccc} \toprule
 (IIB) &0&1&2&3&4&5&6&7&8&9 \\ \midrule
 D($-1$) &&&&&&&&&& \\
 D3 on $\mathcal{S}_\Upsilon$ &--&--&--&--&&&&&& \\
 D3$'$ on $\mathcal{S}_\Gamma$ &&&&&&&--&--&--&-- \\
 D7 &--&--&--&--&&&--&--&--&-- \\\toprule
 \end{tabular*}
 \label{eq:IIBsetup2}
\end{align}
This configuration is known to be the gauge origami in eight dimensions~\cite{Nekrasov:2016ydq}.
The finite quiver case can be understood as the reduction from these affine quiver realizations:
In order to realize the finite quiver $A$ from the affine quiver $\widehat{A}$, we should take the $\mathbb{S}^1$ radius to be infinity in the type IIA setup~\eqref{eq:IIAsetup}.
In the type IIB setup~\eqref{eq:IIBsetup}, on the other hand, this limit corresponds to the limit $\mathbb{S}^1 \to \{\text{pt}\}$ in the Taub--NUT space.

The configuration presented in \eqref{eq:IIBsetup2} gives rise to 4d $\mathcal{N} = 2$ theory.
Further compactifying 45-directions on the curve $\mathcal{C}$, we obtain 5d and 6d theories compactified on a circle and an elliptic curve.
In general, for $\mathcal{C} = \mathbb{C}$ (4d), $\mathbb{C}^\times$ (5d), $\mathcal{E}$ (6d), we have gauge theory with eight supercharges on $\mathcal{S}_{\Upsilon} (\mathcal{S}_{\Gamma}) \times \mathcal{C}^\vee$, which fully uses ten dimensions as $\mathcal{S}_\Upsilon \times \mathcal{S}_\Gamma \times \mathcal{C}$.

\section{Magnificent Four}\label{sec:mag_four}

The construction discussed above has a natural generalization to a complex four-dimensional situation, called the magnificent four~\cite{Nekrasov:2017cih,Nekrasov:2018xsb}, which is formulated as the Donaldson--Thomas invariant of Calabi--Yau four-folds~\cite{Cao:2017swr,Cao:2019tvv}.
From the string theory point of view, it is realized as the D0-D8-$\overline{\text{D8}}$ system (or D$(-1)$-D7-$\overline{\text{D7}}$ system in the IIB theory as in Section~\ref{sec:geometry}).
See, e.g., \cite{Billo:2009di,Billo:2021xzh,Billo:2009gc,Bonelli:2020gku} for related works.

\subsection[C\^4]{$\boldsymbol{\mathbb{C}^4}$}

We consider the four-fold $\mathbb{C}^4$, which associates the fiber of the cotangent bundle with the decomposition,
\[
 \mathbf{Q} = \bigoplus_{i=1}^4 \mathbf{Q}_i , \qquad
 \operatorname{ch} \mathbf{Q}_i = {\rm e}^{\epsilon_i} = q_i .
\]
We apply a similar notation as \eqref{eq:P_Q_notation} with applying the Calabi--Yau condition $\mathbf{Q}_{1234} = 1$.
We define $\mathbf{N} = \mathbf{n} - \bar{\mathbf{n}}$ and $\mathbf{K}$ with the characters
\[
\operatorname{ch} \mathbf{N} = \operatorname{ch} \mathbf{n} - \operatorname{ch} \bar{\mathbf{n}} = \sum_{\alpha=1}^n \big({\rm e}^{a_\alpha} - {\rm e}^{b_\alpha}\big), \qquad
\operatorname{ch} \mathbf{K} = \sum_{k = I}^k {\rm e}^{\phi_I}.
\]
Hence, the framing space $\mathbf{N}$ can be interpreted as the Chan--Paton (super)vector space of D8-$\overline{\text{D8}}$ configuration.
See also~\cite{Kimura:2019msw}.

Defining the observable sheaf
\[
 \mathbf{Y} = \mathbf{N} - \mathbf{P}_{1234} \mathbf{K} ,
\]
we have an analog of the vector multiplet contribution given by
\[
 \mathbf{V} = \frac{\mathbf{Y}^\vee \mathbf{Y}}{\mathbf{P}_{1234}}
 = \mathring{\mathbf{V}} + \mathbf{V}_\text{inst} ,
\]
where we have the perturbative and the instanton parts
\begin{gather*}
 \mathring{\mathbf{V}} = \frac{\mathbf{N}^\vee \mathbf{N}}{\mathbf{P}_{1234}} , \\
 \mathbf{V}_\text{inst} = - \mathbf{N}^\vee \mathbf{K} - \mathbf{Q}_{1234}^\vee \mathbf{K}^\vee \mathbf{N} + \mathbf{P}_{1234}^\vee \mathbf{K}^\vee \mathbf{K}
 \xrightarrow{\mathbf{Q}_{1234} = 1} \mathbf{v}_\text{inst} + \mathbf{v}_\text{inst}^\vee .
\end{gather*}
It has been known that the obstruction theory is not perfect for a smooth four-fold~\cite{Borisov:2017GT}, and hence we should take the ``square root'' contribution to have the virtual tangent bundle
\[
 \mathbf{v}_\text{inst} = - \mathbf{N}^\vee \mathbf{K} + \mathbf{P}_{123}^\vee \mathbf{K}^\vee \mathbf{K} .
\]
We remark that other decomposition with $\mathbf{P}_{234}$, $\mathbf{P}_{134}$, and $\mathbf{P}_{124}$ provides the same result.
Then, the instanton partition function with the topological data $\gamma = (n,k)$ is given by
\[
 Z_\gamma = \mathbb{I}[\mathbf{v}]_\text{inst} = \frac{1}{k!} \frac{[-\epsilon_{12,23,31}]^k}{[-\epsilon_{1,2,3}]^k[-\epsilon_{123}]^k} \oint_{\mathbb{T}_K} \dd{\underline{\phi}} \prod_{I = 1,\ldots, k} \frac{\overline{P}(\phi_I)}{P(\phi_I)} \prod_{1 \le I<J \le k} \frac{[\phi_{IJ} \pm \epsilon_{12,23,31}]}{[\phi_{I} \pm \epsilon_{1,2,3,4}]} [\phi_{IJ}]^2
\]
with the characteristic polynomials, $P(\phi) = P_{\mathbf{n}}(\phi)$ and $\overline{P}(\phi) = P_{\bar{\mathbf{n}}}(\phi)$.
Precisely speaking, we should also take into account the sign factor associated with taking the ``square root'' of the bundle $\mathbf{V}$.
See \cite{Cao:2017swr,Cao:2019tvv, Nekrasov:2017cih,Nekrasov:2018xsb} for detail.
See also \cite{Kanno:2020ybd} for the integral formula.
In this case, the pole and the residue of this integral, namely the equivariant fixed point, is characterized by a~solid partition (four dimensional partition).

\subsection[C\^4/Z\_p]{$\boldsymbol{\mathbb{C}^4/\mathbb{Z}_p}$}

We then consider the four-dimensional orbifold of type $A$, $\mathbb{C}^4/\mathbb{Z}_p$ with $\Upsilon = \mathbb{Z}_p$, where we denote $\Upsilon_0 = \{0,1,\ldots,p-1\}$.
In this case, we consider the following $\mathbb{Z}_p$ action to the coordinate
\[
 \mathbb{Z}_p \colon \ z_i \longrightarrow {\rm e}^{2p_i\pi \iota/p} z_i,
 \qquad i = 1,2,3,4 .
\]
Hence, we have
\[
 \mathbf{Q} = \bigoplus_{i=1}^4 \mathbf{Q}_i \otimes \mathcal{R}_{p_i}
\]
with the Calabi--Yau condition, $\mathbf{Q}_{1234} = 1$ and $\sum_{i=1}^4 p_i \equiv 0$ (mod $p$).
We also denote $p_{ij} = p_i + p_j$, etc.
Using the $\mathbb{Z}_p$ matrix defined in \eqref{eq:Rk_matrix}, we define an analog of the $q$-Cartan matrix as follows:
\[
 \mathbf{C} = \prod_{i=1}^4 (1 - \mathbf{Q}_i \otimes \mathscr{R}_{p_i}) = \mathbf{c} + \mathbf{c}^\dag ,
\]
where we have the half contribution
\begin{gather}
 \mathbf{c}
 = \prod_{i=1}^3 (1 - \mathbf{Q}_i \otimes \mathscr{R}_{p_i})
 = 1 - \sum_{i=1}^3 \mathbf{Q}_i \otimes \mathscr{R}_{p_i} + \sum_{1 \le i < j \le 3} \mathbf{Q}_{ij} \otimes \mathscr{R}_{p_{ij}} - \mathbf{Q}_{123} \otimes \mathscr{R}_{p_{123}} .
 \label{eq:Cartan_4fold}
\end{gather}
Hence, these matrices $\mathbf{C}$ and $\mathbf{c}$ are analogs of $\mathbf{P}_{1234}$ and $\mathbf{P}_{123}$.
We denote the inverse of the $q$-Cartan matrix $\mathbf{C}$ by $\widetilde{\mathbf{C}}$.
Applying the vector notation, $\mathbf{N} = (\mathbf{N}_j)_{j \in \Upsilon_0} = (\mathbf{n}_j - \bar{\mathbf{n}}_j)_{j \in \Upsilon_0}$ and $\mathbf{K} = (\mathbf{K}_j)_{j \in \Upsilon_0}$, and denoting the observable sheaf by
\[
 \mathbf{Y} = \mathbf{N} - \mathbf{C} \mathbf{K} ,
\]
we define the full contribution as follows:
\[
 \mathbf{V} = \mathbf{Y}^\vee \tilde{\mathbf{C}} \mathbf{Y} = \mathring{\mathbf{V}} + \mathbf{V}_\text{inst} .
\]
The perturbative and the instanton parts are given by
\begin{gather*}
 \mathring{\mathbf{V}} = \mathbf{N}^\vee \tilde{\mathbf{C}} \mathbf{N}, \\
 \mathbf{V}_\text{inst} = - \mathbf{N}^\vee \mathbf{K} - \mathbf{K}^\vee \mathbf{N} + \mathbf{K}^\vee \mathbf{C}^\dag \mathbf{K} = \mathbf{v}_\text{inst} + \mathbf{v}_\text{inst}^\vee ,
\end{gather*}
where the half contribution is given by
\[
 \mathbf{v}_\text{inst} = - \mathbf{N}^\vee \mathbf{K} + \mathbf{K}^\vee \mathbf{c}^\dag \mathbf{K} .
\]
Comparing this expression with that for the double quiver theory~\eqref{eq:double_quiver_total}, the $\mathbf{K}^\vee \mathbf{K}$ term becomes identical under identification $\mathbf{c} = \mathbf{c}_\Upsilon \mathbf{c}_\Gamma^+$.
Then, the instanton partition function associated with the topological data $\gamma = (\underline{n},\underline{k}) = \big(n^j,k^j\big)_{j \in \Upsilon_0}$ is given by
\begin{gather}
 Z_\gamma = \frac{1}{\underline{k}!} \oint_{\mathbb{T}_\mathbf{K}} \dd{\underline{\phi}} \prod_{j \in \Upsilon_0} \Bigg[ \prod_{I = 1}^{k^j} \frac{\overline{P}^j\big(\phi_I^j\big)}{P^j\big(\phi_I^j\big)}
 \nonumber \\
 \hphantom{Z_\gamma =}{}
 \times
 \frac{\prod_{I \neq J}^{k^j} \big[\phi_I^j - \phi_J^j\big] \prod_{a<b}^3 \Big( \prod_{J=1,\ldots,k^j}^{I = 1,\ldots,k^{j+p_{ab}}} \big[\phi_I^{j+p_{ab}} -\phi_J^j - \epsilon_{ab}\big] \Big)}{\prod_{a=1}^3 \Big( \prod_{J=1,\ldots,k^j}^{I = 1,\ldots,k^{j+p_{a}}} \big[\phi_I^{j+p_{a}} -\phi_J^j - \epsilon_{a}\big] \Big) \prod_{J=1,\ldots,k^j}^{I = 1,\ldots,k^{j-p_{4}}} \big[\phi_I^{j-p_4} - \phi_J^j + \epsilon_4\big]}
 \Bigg] .
 \label{eq:mag_four_formula_Zp}
\end{gather}
It would be interesting to clarify the residue structure of this integral, which is expected to be characterized using the colored version of solid partitions.
See~\cite{Bonelli:2020gku} for the progress along this direction.

\subsection[C\^4/Upsilon]{$\boldsymbol{\mathbb{C}^4/\Upsilon}$}

In order to consider a generic orbifold $\mathbb{C}^4/\Upsilon$, we should find a $q$-Cartan matrix for the four-fold factorized into that for the three-fold as in~\eqref{eq:Cartan_4fold}.
It is obvious for the Abelian orbifold $\Upsilon = \mathbb{Z}_p$ under the Calabi--Yau condition.
Another possibility is $\big(\mathbb{C}^2 / \Upsilon_A\big) \times \big(\mathbb{C}^2 / \Upsilon_B\big)$ where $\Upsilon_A$ generic, $\Upsilon_B = \mathbb{Z}_p$, and vice versa.
It would be interesting to ask what is a generic form of the factorizable $q$-Cartan matrix.

\section{BPS/CFT correspondence}\label{sec:BPS/CFT_corresp}

The BPS/CFT correspondence is the correspondence between the BPS observables in supersymmetric gauge theory and the vertex operators in CFT-like theory~\cite{Nekrasov:2004UA,Nekrasov:2015wsu}.
An important consequence of this correspondence is the equivalence between 4d gauge theory partition function and a conformal block of the Virasoro/W CFT, a.k.a., AGT-W relation~\cite{Alday:2009aq,Wyllard:2009hg}.
In~\cite{Kimura:2019hnw}, it has been shown that the contour integral form of the instanton partition function also has a conformal block interpretation (free field realization).
See also \cite{Kazakov:1998ji} for an earlier result in this direction.
In this section, we provide the vertex operator construction of the double quiver theory\footnote{We apply the 5d convention in this section.
The 4d and 6d theory can be similarly constructed using the vertex operators discussed in~\cite{Nieri:2019mdl} and~\cite{Kimura:2016dys}.}
and also the magnificent four.

\subsection{Vertex operators}

Denoting the normal ordering symbol by ${:}-{:}$, we define the vertex operators
\begin{alignat*}{3}
& \mathsf{A}_i^j(x) = x^{-\kappa_i^j} \,{ {:} \exp \left( \sum_{n\neq 0} \mathsf{a}_{i,n}^j x^{-n} \right) {:} } , \qquad & &
 \mathsf{Z}_i^j(x) = { {:} \exp \left( \sum_{n\neq 0} \mathsf{z}_{i,n}^j x^{-n} \right) {:} } , & \\
& \mathsf{X}_i^j(x) = { {:} \exp \left( \sum_{n\neq 0} \mathsf{x}_{i,n}^j x^{-n} \right) {:} } , \qquad & &
 \mathsf{Y}_i^j(x) = { {:} \exp \left( \sum_{n\neq 0} \mathsf{y}_{i,n}^j x^{-n} \right) {:} } ,
\end{alignat*}
where each oscillator obeys the commutation relations,
\begin{alignat*}{3}
 & \big[\mathsf{a}_{i,n}^j, \mathsf{a}_{i',m}^{j'}\big] = - \frac{1}{n} c_{\Upsilon,jj'}^{[n]} c_{\Gamma,i'i}^{[n]} \delta_{n+m,0} , \qquad &&
 \big[\mathsf{a}_{i,n}^j, \mathsf{z}_{i',m}^{j'}\big] = \big[\mathsf{x}_{i,n}^j, \mathsf{y}_{i',m}^{j'}\big] = - \frac{1}{n} \delta_{j,j'} \delta_{i,i'} \delta_{n+m,0} , & \\
 & \big[\mathsf{a}_{i,n}^j, \mathsf{x}_{i',m}^{j'}\big] = - \frac{1}{n} \delta_{j,j'} c_{\Gamma,i' i}^{[n]} \delta_{n+m,0} , \qquad &&
 \big[\mathsf{a}_{i,n}^j, \mathsf{y}_{i',m}^{j'}\big] = - \frac{1}{n} c_{\Upsilon,j'j}^{[-n]} \delta_{i,i'} \delta_{n+m,0} , & \\
 & \big[\mathsf{x}_{i,n}^j, \mathsf{x}_{i',m}^{j'}\big] = - \frac{1}{n} \tilde{c}_{\Upsilon,j',j}^{[-n]} c_{\Gamma,i' i}^{[n]} \delta_{n+m,0} , \qquad &&
 \big[\mathsf{y}_{i,n}^j, \mathsf{y}_{i',m}^{j'}\big] = - \frac{1}{n} c_{\Upsilon,j'j}^{[-n]} \tilde{c}_{\Gamma,i' i}^{[n]} \delta_{n+m,0} , & \\
 & \big[\mathsf{x}_{i,n}^j, \mathsf{z}_{i',m}^{j'}\big] = - \frac{1}{n} \tilde{c}_{\Upsilon,j',j}^{[-n]} \delta_{i, i'} \delta_{n+m,0} , \qquad &&
 \big[\mathsf{y}_{i,n}^j, \mathsf{z}_{i',m}^{j'}\big] = - \frac{1}{n} \delta_{j,j'} \tilde{c}_{\Gamma,i' i}^{[n]} \delta_{n+m,0} , & \\
 & \big[\mathsf{z}_{i,n}^j, \mathsf{z}_{i',m}^{j'}\big] = - \frac{1}{n} \tilde{c}_{\Upsilon,j' j}^{[n]} \tilde{c}_{\Gamma,i i'}^{[n]} \delta_{n+m,0} . \qquad &&&
\end{alignat*}
We applied the Adamas operation notation \eqref{eq:Adams_op} to the $q$-Cartan matrices.
We have the following relations among these oscillators:
\[
 \mathsf{a}_{i,n}^j
 = \sum_{k \in \Upsilon_0} \mathsf{x}_{i,n}^k c_{\Upsilon,k j}^{[-n]}
 = \sum_{\ell \in \Gamma_0} \mathsf{y}_{\ell,n}^j c_{\Gamma,\ell i}^{[n]}
 = \sum_{k \in \Upsilon_0, \ell \in \Gamma_0} \mathsf{z}_{\ell,n}^k c_{\Upsilon,kj}^{[-n]} c_{\Gamma,i\ell}^{[-n]} .
\]
These vertex operators are reduced to those discussed in~\cite{Kimura:2019hnw, Kimura:2015rgi}, which were originally motivated by Frenkel--Reshetikhin's construction of $q$-deformation of W-algebras~\cite{Frenkel:1997CMP}:
The $\mathsf{A}$-operator corresponds to the root, while the $\mathsf{X}$- and $\mathsf{Y}$-operators are for the weight in the context of representation theory associated with quiver.

For example, the OPE of $\mathsf{A}$-operators is given by
\[
 \frac{\mathsf{A}_i^j(x) \mathsf{A}_{i'}^{j'}(x')}{{{:} \mathsf{A}_i^j(x) \mathsf{A}_{i'}^{j'}(x') {:}}} = \exp \left( - \sum_{n=1}^\infty \frac{1}{n} c_{\Upsilon,jj'}^{[n]} c_{\Gamma,i'i}^{[n]} \frac{x'^{n}}{x^n} \right) .
\]
Similarly, the OPEs with the operators $\mathsf{X}$ and $\mathsf{Y}$ are given by
\begin{gather*}
 \frac{\mathsf{A}_i^j(x) \mathsf{X}_{i'}^{j'}(x')}{{{:} \mathsf{A}_i^j(x) \mathsf{X}_{i'}^{j'}(x') {:}}} =
 \begin{cases}
 \displaystyle
 \left( 1 - \frac{x'}{x} \right) \left( 1 - \frac{x'}{x} q_{12}^{-1} \right), & i=i', \ j=j',
 \vspace{1mm} \\ \displaystyle
 \left( 1 - \frac{x'}{x} \mu_e^{-1} \right)^{-1}, & e\colon i \to i',\ j=j',
 \vspace{1mm}\\ \displaystyle
 \left( 1 - \frac{x'}{x} \mu_e q_{12}^{-1} \right)^{-1}, & {\rm e}^\vee\colon i' \to i, \ j=j',
 \end{cases}
 \\
 \frac{\mathsf{A}_i^j(x) \mathsf{Y}_{i'}^{j'}(x')}{{{:}\mathsf{A}_i^j(x) \mathsf{Y}_{i'}^{j'}(x'){:}}} =
 \begin{cases}
 \displaystyle
 \left( 1 - \frac{x'}{x} \right) \left( 1 - \frac{x'}{x} q_{12}^{-1} \right), & i=i', \ j=j',
 \vspace{1mm}\\ \displaystyle
 \left( 1 - \frac{x'}{x} \mu_e \right)^{-1}, & {\rm e}^\vee\colon j' \to j, \ i=i',
\vspace{1mm} \\ \displaystyle
 \left( 1 - \frac{x'}{x} \mu_e^{-1} q_{12}^{-1} \right)^{-1}, & e\colon j \to j, \ i=i'.
 \end{cases}
\end{gather*}
Based on these expressions, the instanton partition function of double quiver theory~\eqref{eq:double_quiver_total} with the topological data $\gamma=(n,n',k)$ is written as a correlation function of these vertex operators
\begin{gather}
 Z_\gamma = \frac{\mathscr{Z}_\gamma}{\mathring{Z}} , \qquad
 \mathscr{Z}_\gamma = \frac{1}{\underline{k}!} \oint_{\mathbb{T}_\mathbf{K}} \dd{\underline{x}} \big\langle \mathsf{A}_{\underline{k}}^{-1} \mathsf{X}_{\underline{n}} \mathsf{Y}_{\underline{n}'} \mathsf{Z}_{\underline{n}^\text{f}} \big\rangle , \qquad
 \mathring{Z} = \langle \mathsf{X}_{\underline{n}} \mathsf{Y}_{\underline{n}'} \mathsf{Z}_{\underline{n}^\text{f}} \rangle, \label{eq:Z_correlator_rep}
\end{gather}
where we define the products of the vertex operators as follows:
\begin{alignat*}{3}
 &\mathsf{A}_{\underline{k}} = \prod_{\substack{i \in \Gamma_0 \\ j \in \Upsilon_0}} \prod_{I=1}^{k_i^j} \mathsf{A}_i^j(x_{i,I}^j)
 , \qquad&&
 \mathsf{X}_{\underline{n}} = \prod_{\substack{i \in \Gamma_0 \\ j \in \Upsilon_0}} \prod_{\alpha=1}^{n_i^j} \mathsf{X}_i^j(\nu_{i,\alpha}^j)
 , &\\
 &\mathsf{Y}_{\underline{n}'} = \prod_{\substack{i \in \Gamma_0 \\ j \in \Upsilon_0}} \prod_{\alpha=1}^{n_i^{\prime j}} \mathsf{Y}_i^j(\nu_{\alpha,\alpha}^{\prime j})
 , \qquad&&
 \mathsf{Z}_{\underline{n}^\text{f}} = \prod_{\substack{i \in \Gamma_0 \\ j \in \Upsilon_0}} \prod_{f=1}^{n_i^{\text{f},j}} \mathsf{Z}_i^j(\mu_{i,f}^j) ,&
\end{alignat*}
in the multiplicative notations, $x_{i,I}^j = \exp \big(\phi_{i,I}^j\big)$, $\nu_{i,\alpha}^j = \exp\big(a_{i,\alpha}^j\big)$, $\nu_{i,\alpha}^{\prime j} = \exp \big(a_{i,\alpha}^{\prime j}\big)$, and $\mu_{i,f}^{j} = \exp \big(m_{i,f}^j\big)$.
The integration measure is given by
\[
 \dd{\underline{x}} = \prod_{\substack{i \in \Gamma_0 \\ j \in \Upsilon_0}} \prod_{I=1}^{k_i^j} \frac{\dd{x}_{i,I}^j}{2 \pi \iota x_{i,I}^j} .
\]
From the string theory point of view discussed in Section~\ref{sec:geometry}, the vertex operators, $\mathsf{A}$, $\mathsf{X}$, $\mathsf{Y}$, and~$\mathsf{Z}$, are naturally interpreted as the creation operators of D$(-1)$, D3, D3$'$, and D7 branes.

We have several remarks on the formula.
First, since the expression~\eqref{eq:Z_correlator_rep} is understood with the radial ordering, the dual characteristic polynomials, $\widetilde{P}_i^j$, etc, cannot be directly obtained from this formula.
In order to reproduce them, we should change the ordering, e.g., $\big\langle \mathsf{X}_{\underline{n}} \mathsf{Y}_{\underline{n}'} \mathsf{A}_{\underline{k}}^{-1} \big\rangle$, with tuning the fugacity and the Chern--Simons levels.
See~\cite{Kimura:2019hnw, Kimura:2015rgi} for details.
Second, the normalization factor $\mathring{Z}$ provides the perturbative part of the partition function.
For example, for $(\Upsilon,\Gamma) = \big(\widehat{A}_0,{A}_1\big)$, the $\mathsf{X}$-operator product gives rise to
\begin{align*}
 \frac{\mathsf{X}(\nu) \mathsf{X}(\nu')}{{{:} \mathsf{X}(\nu) \mathsf{X}(\nu') {:}}} & = \exp \left( - \sum_{n=1}^\infty \frac{(1 + q_{12}^{n}) (\nu'/\nu)^n}{n(1-q_1^{n})(1 - q_2^{n})} \right)
 = \Gamma_2\left(\frac{\nu'}{\nu} ;q_1,q_2 \right) \Gamma_2\left(\frac{\nu'}{\nu}q_{12} ;q_1,q_2 \right)
 \nonumber \\
 & = \Gamma_2\left(\frac{\nu'}{\nu} ;q_1,q_2 \right) \Gamma_2\left(\frac{\nu}{\nu'} ;q_1,q_2 \right)
 \Gamma_{\rm e}\left(\frac{\nu}{\nu'} ;q_1,q_2 \right) ,
\end{align*}
where $\Gamma_2(z;q_1,q_2)$ and $\Gamma_{\rm e}(z;q_1,q_2)$ are the $q$-deformed double gamma function~\eqref{eq:mult_gamma} and the elliptic gamma function~\eqref{eq:e_gamma}.
Hence, the $\mathsf{X}$-operator correlator is given by
\[
 \langle \mathsf{X}(\nu_1) \cdots \mathsf{X}(\nu_n) \rangle
 = \prod_{\alpha \neq \beta}^n \Gamma_2\left( \frac{\nu_\beta}{\nu_\alpha}; q_1, q_2 \right)
 \prod_{\alpha < \beta}^n \Gamma_{\rm e}\left( \frac{\nu_\beta}{\nu_\alpha}; q_1, q_2 \right) .
\]
Compared with the expression~\eqref{eq:double_quiver_pert}, the product over the $q$-double gamma function is identified with the perturbative part of the partition function, while the elliptic gamma function part is interpreted as the contribution of the boundary $\partial (\mathbb{C}^2) \times \mathbb{S}^1 = \mathbb{S}^3 \times \mathbb{S}^1$.
We remark that the elliptic gamma function appears in the computation of the superconformal index obtained from the path integral on $\mathbb{S}^3 \times \mathbb{S}^1$.
Similarly, the $\mathsf{Y}$-operator contribution is interpreted as the perturbative contribution associated with the transversal direction.
We also remark that the OPEs between $\mathsf{A}$ and the other operators just provide a rational function, so that the boundary contribution can be absorbed by redefinition of the Chern--Simons level.

Although it is not difficult to write down the generic formula again, we examine the formula~\eqref{eq:Z_correlator_rep} with examples in Section~\ref{sec:CFT_example}.

\subsubsection*{Instanton sum and $\boldsymbol{\mathsf{W}}$-operator}

The total instanton partition function is given by summing over all the topological sectors
\[
 Z = \frac{\mathscr{Z}}{\mathring{Z}} , \qquad
 \mathscr{Z} = \sum_{\underline{k}} \underline{\mathfrak{q}}^{\underline{k}} \mathscr{Z}_\gamma .
\]
In the vertex operator formalism, this summation is organized as follows:
We define the $\mathsf{W}$-operator from the $\mathsf{A}$-operator
\[
 \mathsf{W}_i^j = \oint \frac{\dd{x}}{2 \pi \iota x} \mathsf{A}_i^j(x)^{-1} , \qquad
 \mathsf{W}_{\underline{k}} = \prod_{\substack{i \in \Gamma_0 \\ j \in \Upsilon_0}} (\mathsf{W}_i^j)^{k_i^j} .
\]
This construction is analogous to the screening charge, which is defined by the integral of the screening current.
In fact, the screening current of the associated W-algebra has a close relation to the $\mathsf{A}$-operator~\cite{Frenkel:1997CMP,Kimura:2020jxl}.
The partition function for each topological sector $\gamma$ is given by
\[
 \mathscr{Z}_\gamma = \frac{1}{\underline{k}!} \langle \mathsf{W}_{\underline{k}} \mathsf{X}_{\underline{n}} \mathsf{Y}_{\underline{n}'} \mathsf{Z}_{\underline{n}^\text{f}} \rangle .
\]
This expression is analogous to the Dotsenko--Fateev formula of the conformal block where the $\mathsf{W}$-operator plays a role of the screening charge operator.
The summation over the topological sectors then gives rise to a concise formula for the full partition function
\[
 \mathscr{Z} = \sum_{{k}_i^j = 0}^\infty 
 \prod_{\substack{i \in \Gamma_0 \\ j \in \Upsilon_0}}
 \frac{\big(\mathfrak{q}_i^j\big)^{k_i^j}}{k_i^j !}
 \langle \mathsf{W}_{\underline{k}} \mathsf{X}_{\underline{n}} \mathsf{Y}_{\underline{n}'} \mathsf{Z}_{\underline{n}^\text{f}} \rangle = \big\langle {\rm e}^{\mathsf{W}_\text{tot}} \mathsf{X}_{\underline{n}} \mathsf{Y}_{\underline{n}'} \mathsf{Z}_{\underline{n}^\text{f}} \big\rangle,
\]
where we define $\mathsf{W}_\text{tot} = \sum_{j \in \Upsilon_0}^{i \in \Gamma_0} \mathfrak{q}_i^j \mathsf{W}_i^j$.

\subsection{Examples}\label{sec:CFT_example}

\subsubsection[(A\_1,A\_1) quiver]{$\boldsymbol{(A_1,A_1)}$ quiver}

The simplest example is the finite case, $(\Upsilon,\Gamma) = (A_1,A_1)$.
In this case, the $\mathsf{a}$-oscillator obeys the commutation relation
\[
 [\mathsf{a}_n,\mathsf{a}_m] = - \frac{1}{n} \big(1 + q_{12}^n\big)\big(1+q_{12}^{-n}\big) \delta_{n+m,0} .
\]
Hence, the $\mathsf{A}$ operator product gives rise to
\[
 \prod_{I=1}^k \mathsf{A}(x_I)^{-1} = \prod_{I<J}^k [\phi_{IJ}]^2 [\phi_{IJ} \pm \epsilon_{12}]\, {{:} \prod_{I=1}^k \mathsf{A}(x_I)^{-1} {:}} ,
\]
which reproduces the measure factor in~\eqref{eq:A1A1_inst_part}.
The products with $\mathsf{X}$, $\mathsf{Y}$, and $\mathsf{Z}$ provide the gauge polynomial, the $\mathscr{Y}$-function, and the matter polynomial, respectively.

\subsubsection[(widehat A\_0, widehat A\_0) quiver]{$\boldsymbol{\big(\widehat{A}_0,\widehat{A}_0\big)}$ quiver}

For example, in the case of $(\Upsilon,\Gamma) = \big(\widehat{A}_0,\widehat{A}_0\big)$, we have the commutation relation
\[
 [\mathsf{a}_n,\mathsf{a}_m] = - \frac{1}{n} \big(1 - q_1^n\big) \big(1 - q_2^n\big) \big(1 - q_3^n\big) \big(1 - q_4^n\big) \delta_{n+m,0} ,
\]
which is symmetric among $(q_i)_{i=1,2,3,4}$.
Then, the $\mathsf{A}$ operator product gives rise to
\[
 \prod_{I=1}^k \mathsf{A}(x_I)^{-1}
 =
 \prod_{I<J}^k \frac{[\phi_{IJ}\pm \epsilon_{12,23,31}]}{[\phi_{IJ} \pm \epsilon_{1,2,3,4}]} [\phi_{IJ}]^2 \, {{:} \prod_{I=1}^k \mathsf{A}(x_I)^{-1} {:}} ,
\]
which reproduces the measure factor in~\eqref{eq:epsilon34}.

In this case, the $\mathsf{Z}$-operator product is given by the $q$-quadruple gamma function
\begin{gather*}
 \frac{\mathsf{Z}(\nu) \mathsf{Z}(\nu')}{{{:} \mathsf{Z}(\nu) \mathsf{Z}(\nu') {:}}} = \exp \biggl( - \sum_{n=1}^\infty \frac{(\nu'/\nu)^n}{n\big(1-q_1^{n}\big)\big(1 - q_2^{n}\big)\big(1-q_3^{n}\big)\big(1 - q_4^{n}\big)} \biggr) = \Gamma_4\biggl(\frac{\nu'}{\nu} ;q_1,q_2,q_3,q_4 \biggr) ,
\end{gather*}
which is associated with the perturbative part of the eight-dimensional theory on $\mathbb{C}^4$.

\subsubsection{Magnificent four}

We can similarly consider the partition function of the magnificent four using the vertex operator formalism.
Since it is realized as the D$(-1)$-D7-$\overline{\text{D7}}$ (D0-D8-$\overline{\text{D8}}$) system, we have the following expression:
\[
 {Z}_\gamma = \frac{1}{\underline{k}!} \oint_{\mathbb{T}_\mathbf{K}} \dd{\underline{x}} \big\langle \mathsf{A}_{\underline{k}}^{-1} \mathsf{Z}_{\underline{n}} \big\rangle,
\]
where we modify the $\mathsf{Z}$-operator product
\[
 \mathsf{Z}_{\underline{n}} = {:} \prod_{\substack{i \in \Gamma_0 \\ j \in \Upsilon_0}} \prod_{\alpha=1}^{n_i^{j}} \mathsf{Z}_i^j\big(\nu_{i}^j\big) \mathsf{Z}_i^j\big(\bar{\nu}_{i}^j\big)^{-1} {:} ,
\]
with the multiplicative parameters $\nu_{i,\alpha}^j = \exp\big(a_{i,\alpha}^j\big)$ and $\bar{\nu}_{i,\alpha}^j = \exp\big(b_{i,\alpha}^j\big)$.
Hence, the measure term is the same as before, while the source term is now compatible with the formula~\eqref{eq:mag_four_formula_Zp}.
We remark that this magnificent four partition function is reduced to three-dimensional Donaldson--Thomas theory by tuning the Coulomb moduli~\cite{Nekrasov:2017cih,Nekrasov:2018xsb}, where another type of free field realization is applied~\cite{Benini:2018hjy,Pomoni:2021hkn}.

\appendix
\section[Generic Z\_p orbifold surface]{Generic $\boldsymbol{\mathbb{Z}_p}$ orbifold surface}\label{sec:Cartan_gen}

We consider a complex surface obtained through the resolution of a singularity associated with the following generic $\mathbb{Z}_p$ quotient of $\mathbb{C}^2$:
\begin{align}
 \mathbb{Z}_p \ : \
 \begin{pmatrix}
 z_1 \\ z_2
 \end{pmatrix}
 \ \longrightarrow \
 \begin{pmatrix}
 {\rm e}^{2 r \pi \iota / p} & 0 \\ 0 & {\rm e}^{2 s \pi \iota / p}
 \end{pmatrix}
 \begin{pmatrix}
 z_1 \\ z_2
 \end{pmatrix}
 .
 \label{eq:Zp_rs}
\end{align}
Namely, we consider $\mathbf{Q} = (\mathbf{Q}_1 \otimes \mathcal{R}_{r}) \oplus (\mathbf{Q}_2 \otimes \mathcal{R}_s)$ instead of \eqref{eq:Q_rep}.
This two-dimensional representation can be embedded with SU(2) group only if $r+s \equiv 0$ (mod $p$).
Otherwise we should consider U(2) group embedding.
The case with $r = 1$ corresponds to the Hirzebruch--Jung surface, whose boundary is given by the lens space $L(p;s)$.
We obtain the corresponding $q$-Cartan matrix under the $\mathbb{Z}_p$ action \eqref{eq:Zp_rs} as in \eqref{eq:vect_contribution_upsilon}
\[
 \mathbf{c} = (1 - \mathbf{Q}_1 \otimes \mathscr{R}_{r}) (1 - \mathbf{Q}_2 \otimes \mathscr{R}_{s})
 = 1 - \mathbf{Q}_1 \otimes \mathscr{R}_r - \mathbf{Q}_2 \otimes \mathscr{R}_s + \mathbf{Q}_{12} \otimes \mathscr{R}_{r+s},
\]
and its component is given by
\[
 \mathbf{c}_{ij} = \delta_{i,j} - \mathbf{Q}_1 \delta_{i+r,j} - \mathbf{Q}_2 \delta_{i+s,j} + \mathbf{Q}_{12} \delta_{i+r+s,j} ,
\]
where the indices are understood as mod $p$.
The determinant is given by in the character notation as
\[
 \det c = \big( 1-q_1^{p/\operatorname{gcd}(p,r)} \big)^{\operatorname{gcd}(p,r)} \big( 1-q_2^{p/\operatorname{gcd}(p,s)}\big)^{\operatorname{gcd}(p,s)} ,
\]
where we denote the greatest common divisor of $(x,y)$ by $\operatorname{gcd}(x,y)$ and remark that ${\operatorname{gcd}(p,0) = p}$.
Hence, it is invertible when $q_{1,2} \neq 1$ in general.

\subsection*{Acknowledgement}

I would like to thank Go Noshita and Vasily Pestun for useful discussions.
This work was in part supported by ``Investissements d'Avenir'' program, Project ISITE-BFC (No.~ANR-15-IDEX-0003), EIPHI Graduate School (No. ANR-17-EURE-0002), and Bourgogne-Franche-Comt\'e region.
A part of the results in this paper has been presented at \href{https://indico.ipmu.jp/event/168/}{Representation theory, gauge theory and integrable systems}, February 2018, and \href{https://soochowiashep.github.io/Soochow-first-HEP/}{The 1st SUIAS workshop in HEP: Supersymmetry and Gravitation}, August 2022.
I would like to thank the organizers for invitation and hospitality.
In addition, I am also grateful to the anonymous referees for constructive comments on the manuscript.

\pdfbookmark[1]{References}{ref}
\LastPageEnding

\end{document}